%% file: main.tex
\documentclass[aps,prd,amsmath,amssymb,
twocolumn,
floatfix]
{revtex4-1}
\usepackage[bottom]{footmisc}
\usepackage{graphicx,caption,subcaption}
\usepackage{times}
\usepackage{multirow}
\usepackage{epstopdf}
\usepackage{color}
\usepackage[normalem]{ulem}
\usepackage{amsmath}
\usepackage{lineno}
\graphicspath{{figures/}}

\usepackage[draft, inline, nomargin]{fixme}

\newcommand{\nuebar}{\ensuremath{\overline{\nu}_{e}}}
\newcommand{\nubar}{\ensuremath{\overline{\nu}} }
\newcommand{\uFive}{$^{235}$U}
\newcommand{\uSix}{$^{236}$U}
\newcommand{\uEight}{$^{238}$U}
\newcommand{\pEight}{$^{238}$Pu}
\newcommand{\pNine}{$^{239}$Pu}
\newcommand{\pOne}{$^{241}$Pu}
\newcommand{\npEight}{$^{238}$Np}
\newcommand{\al}{$^{28}$Al}
\newcommand{\cro}{$^{55}$Cr}
\newcommand{\cu}{$^{66}$Cu}
\newcommand{\mn}{$^{56}$Mn}
\newcommand{\he}{$^{6}$He}
\newcommand{\li}{$^{8}$Li}
\newcommand{\va}{$^{52}$V}
\newcommand{\vaFifty}{$^{50}$V}
\newcommand{\vaFiftyOne}{$^{51}$V}

\newcommand{\B}{$\beta^-$}

\renewcommand{\footnote}{\Roman{footnote}}

\begin{document}


\title{Nonfuel Antineutrino Contributions in the High Flux Isotope Reactor}

\input{pfile_author_feb2020.tex}

\collaboration{PROSPECT Collaboration}
\date{\today}
\email{conantaj@ornl.gov}

\begin{abstract}

Reactor neutrino experiments have seen major improvements in precision in recent years. With the experimental uncertainties becoming lower than those from theory, carefully considering all sources of $\overline{\nu}_{e}$ is important when making theoretical predictions. One source of $\overline{\nu}_{e}$ that is often neglected arises from the irradiation of the nonfuel materials in reactors. The $\overline{\nu}_{e}$ rates and energies from these sources vary widely based on the reactor type, configuration, and sampling stage during the reactor cycle and have to be carefully considered for each experiment independently. In this article, we present a formalism for selecting the possible $\overline{\nu}_{e}$ sources arising from the neutron captures on reactor and target materials. We apply this formalism to the High Flux Isotope Reactor (HFIR) at Oak Ridge National Laboratory, the $\overline{\nu}_{e}$ source for the the Precision Reactor Oscillation and Spectrum Measurement (PROSPECT) experiment. Overall, we observe that the nonfuel $\overline{\nu}_{e}$ contributions from HFIR to PROSPECT amount to 1\% above the inverse beta decay threshold with a maximum contribution of 9\% in the 1.8--2.0~MeV range. Nonfuel contributions can be particularly high for research reactors like HFIR because of the choice of structural and reflector material in addition to the intentional irradiation of target material for isotope production. We show that typical commercial pressurized water reactors fueled with low-enriched uranium will have significantly smaller nonfuel $\overline{\nu}_{e}$ contribution. 

\end{abstract}

\maketitle

\tableofcontents

\section{Introduction}


Many experiments have been performed to measure the electron antineutrino (\nuebar{}) flux and spectrum from nuclear reactors over the past several decades to advance our knowledge of the standard model.
Nuclear reactors are intense sources of \nuebar{}; approximately six \nuebar{} per fission are produced, resulting in the emission of $\sim$10$^{20}$ \nuebar{} s$^{-1}$ by a 1~gigawatt electric (GWe) commercial light water reactor. 
Typically, detectors are placed near nuclear reactors to detect \nuebar{} via the inverse beta decay (IBD) reaction. Many experiments have been conducted at commercial nuclear reactors with baselines ranging from tens of meters to hundreds of kilometers. Recent interest in the search for sterile neutrino oscillations has motivated a new series of short-baseline experiments. The need for close proximity to a compact \nuebar{} source and the desire to measure the \nuebar{} production from individual fissile isotopes make research reactors an excellent choice for these experiments~\cite{Heeger:2012tc}. An outline of major neutrino experiments can be found in Ref.~\cite{Kim2016}. 

The detection of \nuebar{} at a nuclear reactor is dependent on many parameters of the reactor and detector systems \cite{DayaBayEvolution}:

\begin{equation}
    \label{eqn:detected_nu_flux}
    \frac{d^2 N\textsubscript{\nuebar{}}(E,t)}{dEdt} = N\textsubscript{prot} \sigma\textsubscript{IBD}(E) \eta \frac{P(E,L)}{4 \pi L^2} \frac{d^2\phi(E,t)}{dEdt} \\,
\end{equation}

\noindent where \textit{N}\textsubscript{\nuebar{}} is the number of neutrinos detected in the active volume, $N$\textsubscript{prot} is the number of target protons in the detector, $\sigma$\textsubscript{IBD} is the energy-dependent IBD cross section, $\eta$ is a detector efficiency parameter, $P$ is the oscillation survival probability, and $L$ is the distance from fission site to IBD interaction position. The last differential term accounts for the magnitude and relative change in the emitted spectrum from the source:

\begin{equation}
    \label{eqn:neutrino_source}
    \frac{d^2\phi(E_{\nuebar},t)}{dE\textsubscript{\nuebar{}}dt}  = \sum_i f_i(t) \frac{dN_i}{dE_{\nuebar}}, 
\end{equation}

\noindent where \textit{f} is the fission rate of isotope \textit{i} and the $dN_i/dE_{\nuebar}$ is the \nuebar{} spectrum of that isotope.

If the focus turns to the reactor as the \nuebar{} source with the fission rate of Equation \ref{eqn:neutrino_source} analogous to power, Equation \ref{eqn:detected_nu_flux} reduces to a simple calculation of the detected \nuebar{} \cite{Klimov1994}:

\begin{equation}
    \label{eqn:simple_neutrino_power}
    \frac{dN_{\bar{\nu}_e}}{dt} = \gamma [1 + k(t)] P\textsubscript{th},
\end{equation}

\noindent where $\gamma$ takes into account all detector properties, $P\textsubscript{th}$ is the thermal power of the reactor, and $k(t)$ takes into account the change in \nuebar{} flux due to isotopic evolution. The Daya Bay \cite{DayaBay,DayaBayEvolution} and RENO \cite{RENO} collaborations have investigated and quantified this $k(t)$ term for commercial reactors. The detected \nuebar{} rate is proportional to reactor power. There exists variation in the signal due to the evolution of isotope fission fractions with fuel burnup, depending on the reactor.

The proportionality of detected \nuebar{} rate to reactor power has been observed in many reactor experiments, most recently Daya Bay \cite{DayaBay,DayaBayEvolution}, Double Chooz \cite{DoubleChooz}, and RENO \cite{RENO}, all of which focused on measuring \nuebar{} disappearance. However, re-evaluation of theoretical predictions \cite{Mention2011,Huber2011,Fallot2012} in preparation for these experiments lead to a 6\% deficit of the observed flux, the ``reactor antineutrino anomaly.'' Additionally, the shape of the overall \nuebar{} spectrum does not agree with predictions with an excess in the 5--7~MeV range, known as the ``bump,'' as the most prominent feature. Causes of these phenomena could be new physics in the form of sterile neutrinos \cite{Mention2011} or incomplete treatment of the complex nature of nuclear reactions and decays in the predictions or both~\cite{Hayes2014,Hayes2015,DwyerLangford2015,Littlejohn2018}.

Previous work has addressed some aspects of reactor design and operation as they affect the reactor \nuebar{} spectrum. The effect of ``nonequilibrium" isotopes, i.e., fission products that have not reached equilibrium contributions, factored in irradiation time as a variable in reactor operation, impacting comparisons with the aggregate beta spectra measured at the Institut Laue-Langevin (ILL) reactor \cite{Sch1985,Kopeikin2003,Mueller2011}. Another observation is the contribution from neutron capture on fission products \cite{Mueller2011,HuberJaffke2015}. Similarly, the contributions from stored spent fuel have been studied \cite{Zhou2012,Brdar2017}. However, the exact contribution of \nuebar{} from nonfission products, primarily via thermal neutron capture on reactor materials, has yet to be addressed. Huber and Jaffke discussed the \nuebar{} contributions on nonequilibrium isotopes, but this effect was examined for neutron capture on fission products only \cite{HuberJaffke2015}. These contributions can be additional terms in Equation~\ref{eqn:neutrino_source}:

\begin{multline} \label{eqn:neutrino_source_corrections}
    \frac{d^2\phi(E_{\nuebar},t)}{dE_{\nuebar}dt}  = \sum_i f_i(t) \frac{dN_i}{dE_{\nuebar}} c_i\textsuperscript{ne}(E_{\nuebar},t) \\ + s\textsubscript{SNF}(E_{\nuebar},t) + a\textsubscript{NF}(E_{\nuebar},t),
\end{multline}

\noindent where the contribution from nonequilibrium isotopes, $c_i\textsuperscript{ne}$, is a correction factor to the isotope spectra. The contribution from spent nuclear fuel, $s\textsubscript{SNF}$, and the nonfuel activations, $a_{NF}$, are additional terms. Note that all these additional contributions are highly reactor-specific and time dependent. Most modern reactor \nuebar{} experiments account for the time dependent $c_i\textsuperscript{ne}$ and $s\textsubscript{SNF}$ but do not account for $a\textsubscript{NF}$ because it has been underexplored or considered to be a trivial contribution \cite{Hayes2015}.


The goal of this paper is to develop a formalism for determining the nonfuel candidates that produce \nuebar{} above the IBD threshold. The Precision Reactor Oscillation and Spectrum Measurement (PROSPECT) experiment at the High Flux Isotope Reactor (HFIR) at Oak Ridge National Laboratory (ORNL) is used as a case study to apply this formalism. The unique configuration of this research reactor provides an opportunity to highlight materials and processes that can make non-negligible contributions to the total emitted \nuebar{} spectrum. Research reactors typically have very different design and missions compared to commercial nuclear reactors, which results in significantly different nonfuel contributions. The formalism defined here can be used by all the reactor \nuebar{} experiments, but it is particularly important for the experiments using research reactors like PROSPECT~\cite{PROSPECTdetector}, STEREO~\cite{STEREO}, and SoLid~\cite{SoLid} to account for the nonfuel contributions in their predictions.

There has been increased interest in measuring the coherent elastic neutrino-nucleus scattering (CE$\nu$NS) reaction using reactors as a source after a first measurement of this reaction by the COHERENT experiment \cite{COHERENT}. Using reactors as the source measuring the CE$\nu$NS requires an update in the predicted \nuebar{} spectrum below the IBD threshold, which has not been given much attention so far \cite{Chooz2019}. Because nonfuel sources of \nuebar{} primarily contribute at low energies, including these contributions in the \nuebar{} predictions is critical. The methodology provided here can be individually used by each experiment to predict the \nuebar{} spectrum provided by the reactor.

Section \ref{sec:selection} outlines the methodology for selection of candidates for nonfuel \nuebar{} emissions in a nuclear reactor, and this methodology can be applied to any reactor. Section~\ref{sec:hfir_case_study} presents HFIR as a case study for the selection process. A list of candidate isotopes are considered for HFIR, and the details of those isotopes in the reactor are discussed with the materials grouped into three general categories: structural, reflector, and target. Section~\ref{sec:reactor_modeling} discusses the reactor modeling methodology and its uncertainty considerations. In Section~\ref{sec:nu_conversion}, reactor modeling to obtain reaction rates and conversion to \nuebar{} spectrum are performed. Section~\ref{sec:lwr_comparison} discusses the extension of this work to commercial nuclear power plants, and Section~\ref{sec:conclusions} contains the conclusions. Finally, Section~\ref{sec:acronyms} contains a list of relevant acronyms.



\section{Nonfuel Production of Antineutrinos}
\label{sec:selection}

This section discusses a procedure for selection of nonfuel \nuebar{} sources in a reactor. One feature of reactor \nuebar{} sources that has sometimes been neglected is the emission from nonfuel materials. The design and operation of certain reactors require specific considerations for nonfuel materials. Nonfission reactions, such as neutron capture, can generate beta-decaying products that are accompanied by \nuebar{}. These \nuebar{} from fission sources are an additional contribution that must be considered for precision measurements at certain classes of reactors. The contribution of these nonfuel sources needs to be taken into account in \nuebar{} predictions.

The isotopes of most concern for predicting an accurate fission \nuebar{} spectrum would be those that contribute to the \nuebar{} flux coming from the core materials, which alters the \nuebar{} fission spectrum. Neutron capture reactions---such as (n,$\gamma$), (n,2n), or (n,p)---release significantly less energy than fission reactions; therefore they contribute negligibly to the core power. \nuebar{} production that is not tracked via the power level disrupts the predicted linear relationship between detected \nuebar{} and power level \cite{Bowden2009}. The isotope content of all nonfuel materials in or near the reactor core must be evaluated for their ability to produce \nuebar{}. The isotopes with significant contributions will be referred to as ``antineutrino candidates.'' Here, reaction rates greater than 0.1\% of the fission rate are considered significant.

To contribute significantly to the spectrum, the combination of parent and daughter isotopes of the neutron capture reaction must fulfill certain criteria. These criteria serve to identify potential contributors to the antineutrino flux. Each isotope needs to fulfill all criteria, but in some cases unmet criterion are balanced by enhancements in other criteria.


First, an antineutrino candidate must have a relatively high concentration in the core. It cannot be contained in trace amounts or be infrequently irradiated in the core. This criterion ensures a sufficient number of target atoms for neutron capture. In addition, a high abundance in the core relative to other isotopes is ideal to maximize the number of target atoms. No quantitative criteria is given here. Although commercial reactors have a smaller number of isotopes present in the core, many isotopes can be considered for research reactors.

Second, the neutron-induced reaction of interest must have a non-negligible neutron cross section to produce the daughter. The reaction of interest is almost always neutron capture (i.e., $^A_Z$X$+ ^1_0$n$ \rightarrow ^{A+1}_Z$ X followed by $\gamma$ emission), but reactions that result in the ejection of other particles (e.g, $\alpha$, $^3$H, etc.) also can result in daughter isotopes prone to \B{} decay. Because the neutron-induced reaction rate of an isotope $i$ ($R_i$) is a product of the parent isotope concentration (\textit{N$\textsubscript{p}$}) and energy-dependent neutron cross section ($\sigma_i$) and neutron flux ($\phi$), the second criteria seeks to have a maximum of this product:

\begin{equation}
R_i = N\textsubscript{p}(t) \int \int \phi(\vec{r},E,t) \sigma_{i}(\vec{r},E) dE d\Vec{r}
\label{eqn:rr}
\end{equation}

\noindent For example, a structural material has a relatively high atomic concentration in the core but a relatively low cross section, whereas neutron poisons for reactivity control have the inverse characteristic. Both of these can still be considered as \nuebar{} candidates due to the product of concentration and cross sections. In this work, a non-negligible cross section is defined to be greater than 0.1~barns, although some exceptions are made because of the combination of criteria one and two. Because most \nuebar{} experiments have occurred at thermal reactors, thermal neutron-induced reaction are primarily considered here. The mean energy of thermal neutrons in a nuclear reactor is 0.0253~eV, for which ENDF/V-VII.1 cross sections can be readily obtained \cite{ENDF7}. Fast neutron-induced reactions can be important in certain areas of the core or for fast neutron spectrum reactors, which is beyond the scope of this work.


Third, the daughter product ($^{A+1}_ZX$) must $\beta^-$ decay with a short half-life relative to the cycle of the reactor so that it is generated with a sufficient activity. If the half-life is too long relative to the reactor cycle length, it will not decay with a high enough frequency. The relative magnitude of half-life to cycle length will determine how quickly, if at all, the activity will reach secular equilibrium with its production rate. Half-lives of up to a certain length relative to the cycle length may be considered depending on the application. For example, isotopes with a half-life two orders of magnitude lower than the cycle length will have their activity saturated for 95\% of the cycle. This value will be used for this criterion, although most activated products have half-lives much shorter than this. A short time until an isotope reaches its saturated activity increases the value and decreases the time variation of the \nuebar{} contribution.

Fourth, the $\beta^-$ transition of the daughter must release enough energy to be above the IBD threshold of 1.8~MeV ($E_{\nuebar{},max} = Q - E_{\gamma} >$ 1.8~MeV). The energy released and final state energy can be retrieved from the Evaluated Nuclear Structure File (ENSDF) database \cite{ENSDF} maintained by the National Nuclear Data Center. The IBD threshold requirement is applied to all beta branches of each isotope taking into account the individual abundances. This paper focuses on contributions above the 1.8~MeV IBD threshold; evaluation of nonfuel contributions for detectors sensitive to other \nuebar{} reactions should take into account a lower energy threshold, as in Ref.~\cite{Deniz2010}.


An isotope that fulfills all of these criteria is considered as an antineutrino candidate. In a reactor, the concentration of the candidate, $N_i$, from its parent, $N\textsubscript{p}$, assumed to be stable and not appreciably burnt out, is equivalent to:

\begin{equation}
N_i(t) = \frac{N\textsubscript{p}(t)}{\lambda_i} \left[ 1 - e^{-\lambda_i t} \right] \int \phi(E,t) \sigma_{i}(E) dE
\label{eqn:concentration_activity},
\end{equation}

\noindent where $\lambda_i$ is the decay constant of the daughter isotope ($\lambda = \textnormal{ln} 2 / t_{1/2}$). If the decay constant of the product is large (meaning a short half-life) relative to the irradiation period, the decay term quickly declines and the candidate concentration becomes proportional to the time-dependent neutron flux and parent isotope concentration.


For any reactor \nuebar{} experiment, the above criteria can be applied to its reactor materials to select non-fissionable isotopes that may contribute significantly to the reactor \nuebar{} spectrum. The selection process will result in isotopes that should be considered for reactor analysis and modeling to quantify nonfuel \nuebar{} rates. Each reactor can be analyzed based on the materials under consideration.

\section{Case Study: HFIR}
\label{sec:hfir_case_study}

For this paper, HFIR is used as a case study. As a research reactor, HFIR is smaller than traditional commercial reactors and is not used to generate electricity. It also is fueled with highly enriched uranium, whereas typical commercial reactors are fueled with low-enriched uranium. HFIR currently hosts the PROSPECT detector, which is measuring the \nuebar{} flux from HFIR. HFIR is similar in design to other research reactors, such as the National Bureau of Standards Reactor \cite{Cheng2004,Diamond2014}, the ILL reactor \cite{ILL} which hosts the STEREO experiment \cite{STEREO}, and the BR2 reactor in Belgium \cite{BR2} which hosts the SoLid experiment \cite{SoLid}. The study of the nonfuel \nuebar{} could be applicable to these other highly enriched uranium  reactors.

\begin{figure}
\centering
\includegraphics[width=0.45\textwidth]{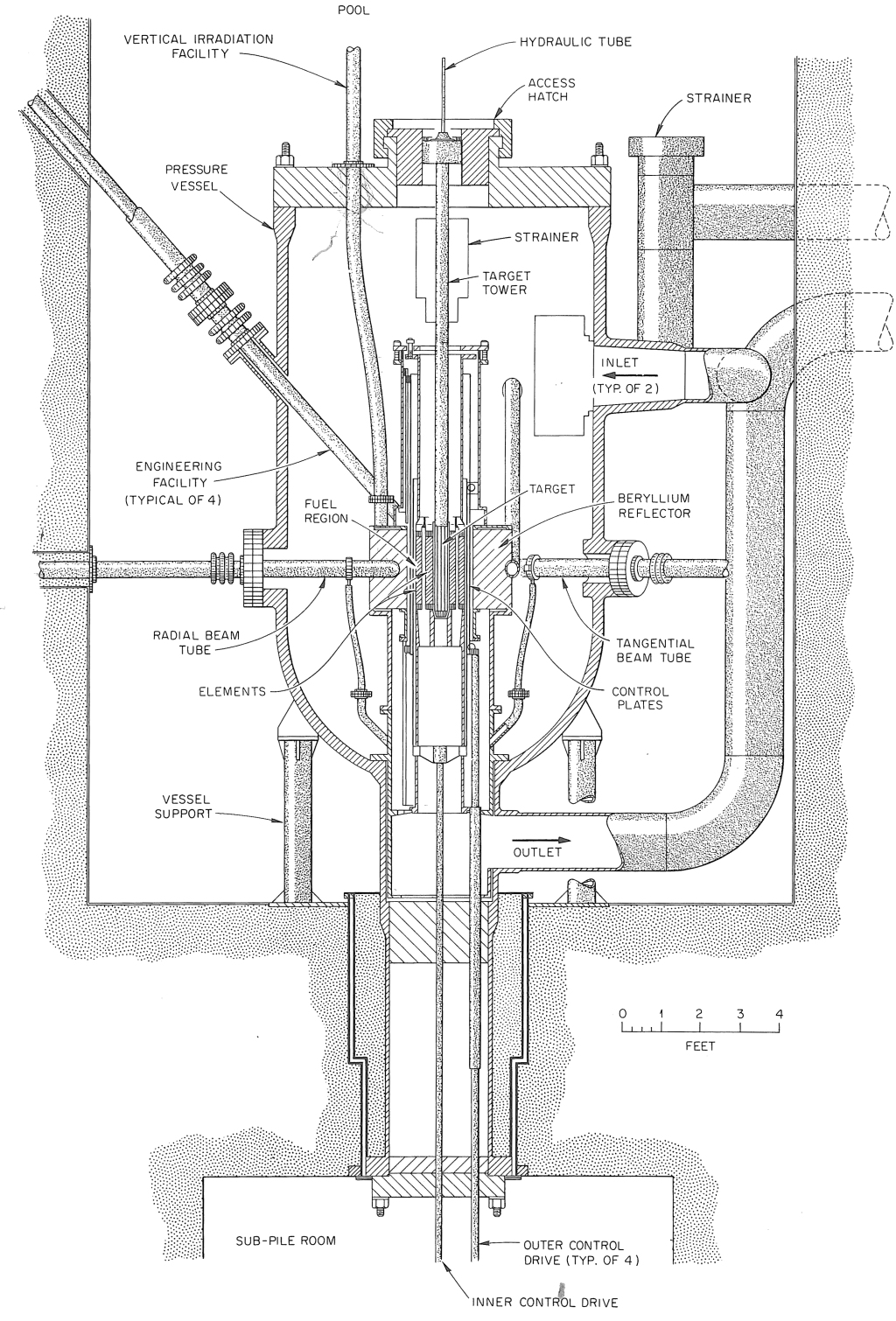}
\includegraphics[width=0.45\textwidth]{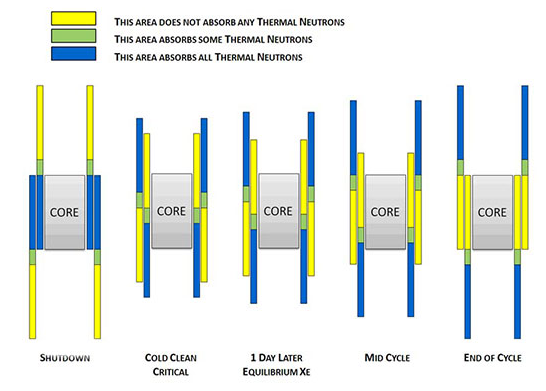}
\caption{Side view of HFIR with core regions (top) and movement of inner and outer CEs throughout the cycle (bottom).}
\label{fig:hfir_regions}
\end{figure}

HFIR is a major U.S. research reactor with missions of neutron scattering, isotope production, materials irradiation, and neutron activation analysis \cite{Cheverton}. It is one of the few highly enriched uranium--fueled research reactors in the United States and has been operating since 1965. HFIR is a compact reactor that can attain high thermal neutron fluxes---greater than $2 \times 10^{15}$ cm$^{-2}$ s$^{-1}$---in its central region. It nominally operates at a power of 85 megawatts thermal (MWt) for a cycle length of 23--26~days, i.e., 1,955--2,210 megawatt days (MWd) of operation with seven cycles annually. Figure \ref{fig:hfir_regions} shows the side view of HFIR.

The central region of the core is the flux trap target (FTT) region. The FTT region contains a total of 37 target positions, which includes 30 interior positions, 6 peripheral target positions, and one hydraulic tube. The contents of the FTT vary from cycle to cycle depending on experimental demand for isotope production and materials irradiation. A model with a representative loading, for example, contains target materials composed of V, Ni ($^{62}$Ni), Mo, W, Se, Ni, Fe, and Cm~\cite{Chandler2016}. The curium targets are used to produce $^{252}$Cf \cite{Hogle2012}, which results in its spontaneous fissions and other neutron-induced fission of higher actinides. In more recent cycles since that report, experiments have included previously mentioned isotopes as well as silicon carbide, steels, and other ferritic alloys. These isotopes are important for PROSPECT, which was deployed in early 2018.

Radially outward of the FTT are the two fuel element regions, the inner and outer fuel elements (IFE/OFE). The fuel is a U$_3$O$_8$-Al dispersion fuel (uranium dispersed in an aluminum matrix) enriched to 93\% by mass \uFive{} (5--6\% \uEight{} and ~1\% \uSix{}) and manufactured in the form of involute plates \cite{Knight1968}. The fuel region is contoured along the arc of the involute to allow for sufficient thermal safety margin. The IFE contains a burnable poison, $^{10}$B, to flatten the power distribution and ensure a longer cycle. The IFE and OFE contain 171 and 369 fuel plates, respectively, and have separate fissile loadings. Fresh IFE and OFE fuel assemblies are loaded into the core for each cycle, unlike most commercial reactors that operate with some previously irradiated fuel elements containing plutonium.

The fuel regions are surrounded by two concentric control elements (CEs). Both control elements are partially inserted at the beginning of cycle (BOC) and are gradually withdrawn in opposite directions throughout the cycle. The inner control element (ICE) is the control cylinder that descends throughout the cycle; the outer control element (OCE) is a set of four safety plates, each of which can individually scram the reactor, move upward throughout the cycle. The CE positions at various points in the cycle are shown in Figure~\ref{fig:hfir_regions}. Both control elements contain Eu, Ta, and Al in their absorbing regions \cite{Chandler2016}. The end of cycle (EOC) occurs when both elements are fully withdrawn and the reactor can no longer maintain criticality. Both the ICE and OCE are replaced approximately every 50 cycles.

The beryllium reflector occupies the outermost radial region and serves to moderate neutrons for reflection back into the active core or transport them down the beam tubes. The reflector region is split up into three regions: the removable (RB), semi-permanent (SPB), and permanent (PB) beryllium regions. The RB is replaced every several years, and the SPB and PB are replaced every few decades. The PB contains 22 vertical experimental facilities (VXFs), including inner small, outer small, and large VXFs. The four horizontal beam tubes (HBs) penetrate the outer radial areas in order to support cold and thermal scattering experiments. Recent cycles have included neptunium oxide (NpO$_2$) targets to produce $^{238}$Pu for the National Aeronautics and Space Administration (NASA) \cite{Chandler2015,Chandler2016physor,Hurt2016}. All materials in the various components of the reflector regions are included in the analysis. Because reflector regions are exposed to neutron flux, they build up substantial neutron poisons, primarily $^6$Li and $^3$He, over multiple irradiation cycles. Several reactions that produce these poisons and \nuebar{} candidates rely on fast neutrons, whereas captures in the structural and target materials occur mostly from thermal neutrons.


The \nuebar{} candidate selection process of Section~\ref{sec:selection} is applied to the nonfuel materials in HFIR. The reactor is first divided into different regions according to primary function. Then, a mix of publicly available and internal data at HFIR is used to determine average quantities of materials in the core during a typical cycle. The composition of the fuel elements is well documented and outlined in Ref.~\cite{Chandler2016}. The control element (CE) and reflector materials change in composition with increasing irradiation time in the reactor. The target materials have the potential to change each cycle according to user demand; a representative loading of targets in recent cycles is outlined in Ref.~\cite{Chandler2016}. Isotopic constituents of these materials are analyzed according to the four step selection process to generate \nuebar{} candidates to be analyzed with reactor modeling. Candidates with contributions of greater than 0.1\% are considered because this is the typical statistical uncertainty in reaction rate calculations.


Antineutrino candidate selection process results can be seen in Table \ref{tab:antinu_candidates}. The $\beta^-$ decays of antineutrino candidates that are to be considered include three main groups. The first is structural materials, which includes \al{}, \cro{}, \cu{}, and \mn{}. The second is the beryllium reflector, which includes \he{} and \li{}. The last is the target materials, which include \va{} in the FTT and two actinide targets, curium in the FTT, and neptunium in the VXFs. The next step for the antineutrino candidates is to quantify their activities in the reactor, discussed in Section \ref{sec:reactor_modeling}, and convert the activities to \nuebar{} spectra to compare with the nominal reactor spectrum in Sections~\ref{sec:nu_conversion}--\ref{sec:conclusions}.

\begin{table*}
\caption{A summary list of the antineutrino candidates in HFIR. The isotopes are grouped in coarse categories according to function or region in the reactor. The requirements for \nuebar{} candidate selection previously described are in bold in the second row of the table. The failed criteria for each isotope, if any, is listed in the right-most column.}
    \begin{tabular}{|ccc|c|c|c|c|c|c|c|c|c|}
    \hline
   &  El. & A & Abundance & ENDF/B-VII.1 & Daughter & t$_{1/2}$ & Q & E$_{final}$ & E$_{\beta,max}$ (MeV) & Criteria \\
   &  & & (\%) & $\sigma$ (barns) &  & (s) & (MeV) & (MeV) & [Mutliple] & Failed \\
    \hline
    \textbf{Requirement} & & & \textbf{1} & \textbf{2} & \multicolumn{2}{|c|}{\textbf{3}} & \multicolumn{4}{|c|}{\textbf{4}} \\
      &  &   & \textbf{High} & \textbf{High} & \textbf{$\beta^-$decay} & \textbf{Low} & \textbf{High}  &       & \textbf{$>$ 1.8 MeV} &  \\
    \hline
  Structural &  Al    & 27 & 100.0 & 0.23  & $^{28}$Al & 1.34E+02 & 4.64  & 1.78  & 2.86 & None \\
   &  Fe    & 54 & 2.8   & 2.25  & $^{55}$Fe &  &       &       & & 3   \\
   &       & 56 & 91.8  & 2.59  & $^{56}$Fe &  &       &       & & 3  \\
   &       & 57 & 2.1   & 2.43  & $^{58}$Fe &  &       &       & & 3  \\
   &       & 58 & 0.3   & 1.00  & $^{59}$Fe & 3.84E+06 & 1.57  &      &  & 4  \\
   & Cr    & 50 & 4.3   & 15.40 & $^{51}$Cr &  &       &       & & 3  \\
   &       & 52 & 83.8  & 0.86  & $^{53}$Cr &  &       &       & & 3  \\
  &        & 53 & 9.5   & 18.09 & $^{54}$Cr &  &       &       & & 3   \\
   &       & 54 & 2.4   & 0.41  & $^{55}$Cr & 2.10E+02 & 2.60  & 0.00  & 2.60 & None  \\
   & Cu    & 63 & 69.2  & 4.47  & $^{64}$Cu &  &       &       & & 3   \\
   &       & 65 & 30.8  & 2.15  & $^{66}$Cu & 3.07E+02 & 2.64  & 0.00  & 2.64 & None  \\
   & Mg    & 24 & 79.0  & 0.05  & $^{25}$Mg &  &       &       & & 3    \\
  &        & 25 & 10.0  & 0.19  & $^{26}$Mg &  &       &       & & 3   \\
  &        & 26 & 11.0  & 0.19  & $^{27}$Mg & 5.73E+02 & 2.61  & 0.84  & 1.77 & 4 \\
   & Mn    & 55 & 100.0 & 13.27 & $^{56}$Mn & 9.28E+03 & 3.70  & Various & [0.250,2.849] & None   \\
  \hline
  Reflector & Be    & 9 & 100.0 & 0.04$^a$ & $^{6}$He & 8.07E-01 & 3.50  & 0.00 & 3.50 & None   \\
            &      & 10 & trace &  & $^{11}$B & 4.75E+13 & 0.55  &  & & 1,2   \\
            & Li    & 7 & 92.41$^b$ & 0.04 & $^{8}$Li & 8.40E-01 & 16.00  & 3.03 & 12.97 & None  \\
  \hline
 Poisons &  B     & 10 & 80.1  & 3842.56 & $^{7}$Li &  &       &   & & 3      \\
 and CEs &        & 11 & 19.9  & 0.01  &  &       &       &       & & 3 \\
  &  Eu    & 151 & 47.8  & 9200.73 & $^{152}$Eu & 4.22E+08 &  &  & & 3    \\
  &        & 153 & 52.2  & 358.00 & $^{154}$Eu & 2.71E+08 &     &       & & 3  \\
  &  Nb    & 93  & 7.59  & 1.16  & $^{94}$Nb & 6.41E+11 & 2.04  &       & & 3   \\
  &  Ta    & 181 & 99.99 & 8250.44  & $^{182}$Ta & 9.89E+06 & 1.81  & Various  &  & 4   \\
  \hline
  Targets & V     & 51 & 99.75 & 4.92  & $^{52}$V & 2.25E+02 & 3.97  & 1.434 & 2.54 & None \\
  (FTT + &  Mo    & 98 & 24.4  & 0.13  & $^{99}$Mo & 2.38E+05 & 1.36  &      & & 4   \\
  VXFs)&  Se    & 78 & 23.8  & 0.43  & $^{79}$Se &  &       &       & & 3  \\
  &        & 80 & 49.6  & 0.61  & $^{81}$Se & 1.11E+03 & 1.59  &      & & 4   \\
  &  Ni    & 58 & 68.1  & 4.22  & $^{59}$Ni &  &       &  & & 3     \\
   & Np    & Various &       &       &  Various$^c$     &       & & None       & & Various$^c$ \\
   & Cm    & Various &       &       &  Various$^c$     &       & & None       & & Various$^c$ \\
    \hline
    \end{tabular}%
    
$^a$The cross-section listed is in the fast region due to the high energy threshold $^9$Be(n,$\alpha$) reaction \\
$^b$With a fresh beryllium reflector, no $^7$Li is present but it is produced gradually in its lifetime \\
$^c$Np, Cm, and products to which they transmute are fissile and produce fission \nuebar spectra that have been relatively unexplored  \\
  \label{tab:antinu_candidates}%
\end{table*}%

\section{Reactor Modeling and Simulation}
\label{sec:reactor_modeling}

After identifying antineutrino candidates for HFIR, the next step is to quantify the neutron-induced reaction rates in a typical cycle of the reactor. The modeling methodology is to build on a HFIR computer model developed by ORNL staff~\cite{Chandler2016} using the Monte Carlo particle transport code MCNP \cite{MCNP5,MCNP6}. This model includes information and advancements from a HFIR Cycle 400 model \cite{Xoubi2004,Ilas2015}, including explicit modeling of the fuel plates and a representative target loading, and is the basis for neutronic safety and performance calculations at HFIR. Models exist for BOC and EOC as well as in single day time steps for each day of the cycle; the isotopics for each day are calculated from the VESTA depletion code \cite{VESTA}.

Reaction rate calculations are added in MCNP to obtain the energy-dependent neutron flux and reaction rates in user-defined, discrete cells containing the isotope of interest, and phantom materials (described in Ref~\cite{MCNP6}) are added to obtain isotope-dependent reaction rates. The lack of phantom materials in a tally results in total reaction rates in a cell (e.g. for fission rates in a fuel cell summed over those for \uFive{}, \uEight{}, \pNine{}, \pOne{}). MCNP cells are user-defined according to regions bound by surface descriptions (e.g., planes, spheres, cylinders). Volumes of these cells range from less than 1~cm$^3$ for fuel and some flux trap cells to hundreds of cubic centimeters for reflector regions. Tally results in MCNP are reported per unit source particle (e.g., neutron). To normalize to absolute rates for comparison with fission rates, the power normalization factor (PNF), expressed in terms of a neutron rate in units seconds$^{-1}$, sometimes called the source term \textit{S}, \cite{Snoj2006,Sterbentz2013} is used:

\begin{equation}
PNF = S = \frac{P\textsubscript{th} \nu}{k\textsubscript{eff} Q\textsubscript{fiss}},
\label{eqn:pnf}
\end{equation}

\noindent where \textit{P} is the thermal power of the reactor, $\nu$ is the number of neutrons generated per fission, $k\textsubscript{eff}$ is the criticality eigenvalue reported in MCNP, and $Q\textsubscript{fiss}$ is the energy released per fission. Typical values for $\nu$ are 2.4 for \uFive{} and 2.9 for \pNine{}. $k\textsubscript{eff}$ is unity for a critical reactor. The $Q\textsubscript{fiss}$ is close to 200~MeV for uranium and plutonium isotopes \cite{Kopeikin2012,Ma2013}. The PNF in HFIR MCNP simulations is assumed to be accurate because the models result in eigenvalues close to unity (with small statistical error) and the energy dependence of $\nu$ and $Q\textsubscript{fiss}$ are negligible for each fissile isotope~ \cite{Zerovnik2014}. Owing to the constant power and little fuel evolution, the PNF changes by 0.1\% throughout a cycle and is therefore considered to be constant.

The goal is to calculate the core reaction rate $R\textsubscript{core}$ for each candidate for each isotope $i$ for each cell $j$ in the model, combining Equation~\ref{eqn:rr} in discretized form and Equation~\ref{eqn:pnf}:

\begin{equation}
R_{i}\textsubscript{,core}(t) = \frac{P\textsubscript{th} \nu}{k\textsubscript{eff} Q\textsubscript{fiss}} \sum_{j=1}^{M\textsubscript{cells}} \frac{N\textsubscript{p,}_j(t)}{\lambda_i} \int_{E} \phi_j(E,t) \sigma_{i}(E) dE
\label{eqn:reaction_rate}
\end{equation}

\noindent so the atomic concentration of the isotope $N_i$ in the cell $M_i$ is multiplied by the integral, i.e., the output of the MCNP reaction rate tally. Given a low half-life of the product and low time variation of the reaction, the core production rate is approximately equal to its activity early into the cycle, i.e.,

\begin{equation}
A_i \approx{} R_{i}\textsubscript{,core}
\label{eqn:act_rr}
\end{equation}

If not replaced every cycle, some of the \nuebar{} candidates evolve in concentration throughout several cycles. When necessary, the COUPLE and ORIGEN (Oak Ridge Isotope Generation) modules in the SCALE modeling and simulation suite~\cite{SCALE62} are used for production, depletion, and decay of these isotopes. The COUPLE and ORIGEN sequences are also used for other isotopes as a cross-check to verify constant concentration ($dN_i/dt \approx 0$) within the duration of the cycle. For COUPLE/ORIGEN, an energy group-dependent neutron flux, total flux, and BOC cell isotope concentrations are required inputs. These inputs are obtained from the MCNP outputs, which are generated for each day in the cycle. The MCNP cases provide the group-spectra using a 44-group energy structure, a collapsed version of the commonly used 238-group structure used in neutron activation problems \cite{SCALE62}. Therefore, the MCNP stand-alone and MCNP combined with ORIGEN inputs are not expected to differ substantially unless the parent isotope has undergone significant transmutation. Note that MCNP uses continuous energy cross sections based on ENDF whereas the multi-group COUPLE/ORIGEN approach was based on using the MCNP binned flux spectrum and JEFF for generating one-group cross sections.

The missions, design, and operation of HFIR allow for a large number of materials to be present and irradiated during a given cycle. In searching for candidate isotopes that could contribute to the \nuebar{} spectrum, all areas of the reactor discussed previously were considered. This includes isotopes in the materials that make up the structural, control element, and reflector regions in addition to the large variety of target materials that can be in the FTT positions or VXFs in the reflector region.

The modeling and simulation provide high-precision calculations of the isotope-dependent fission rates in the core. The fission rate changes negligibly from $2.64 \times 10^{18}$ to $2.65 \times 10^{18}$ s$^{-1}$ from BOC to EOC due to the evolution of the power distribution and gamma radiation. The fission fraction of \uFive{} remains above 99.5\% throughout the cycle. The fission rate is important in determining the \nuebar{} production from fission versus \nuebar{} candidates.

The uncertainty in such reactor model predictions arises from a variety of components. These include, but are not limited to, the uncertainty in (1) model creation such as the precision to which geometry and material compositions are known, (2) nuclear data, and (3) the modeling methodology itself. In the first case, HFIR has a consistent loading except for target and reflector compositions, which change from cycle to cycle. The variation is expected to be small becasue of consistent fuel loading and power distribution within the core. Previous analysis specific to HFIR have found that geometries and isotope concentrations of reactor components agree well with engineering drawings and material specifications, but note that the impurity levels among fabrications may vary which can result in changes in isotope concentrations in components in reality compared to those modeled~\cite{Peplow2004,Cheverton}. Note, the model detail level is higher in the fuel and near experiments of interest as these calculations served to provide precise neutron flux values. Therefore the uncertainty associated with model isotope concentrations is assumed to be $\leq$ 1\%. Regarding nuclear data, for most reactions induced by thermal neutrons, the uncertainties in cross sections vary from 0.1\% to 0.5\% for well-known isotopes and several percent for isotopes with more uncertain cross sections. Isotope concentration uncertainties, and thereby reaction rates, in time-dependent calculations can be low for actinides but tens of percent for some fission products \cite{Diez2015}. Most of the candidate isotopes in this study have uncertainties on the lower side of that range. The third type is the uncertainty from the methodology itself. In the past two decades, Monte Carlo codes have been increasingly used for neutron transport calculations, and the uncertainty associated with the methodology itself is largely statistical. With recent improvements in computational power during the past decade, statistical uncertainties can be obtained at the subpercent level for flux and reaction rates. In codes predominantly used for reactor analysis, such as MCNP and SCALE, isotope concentration uncertainties have been validated to several percent or better for many benchmarking problems \cite{MCNPvalidation,Ilas2010}. In short, the propagation of uncertainty in reactor simulations is not straightforward and has several considerations. However, the uncertainty associated with reactor calculations is expected to be tenths of a percent to several percent depending on the isotope. In addition, larger uncertainties exist outside the scope of reactor simulation uncertainties, such as the precision of the reactor power level \cite{Conant2019,Radaideh2018}.

\section{Calculation of Nonfuel Excess in \nuebar{} Spectrum}
\label{sec:nu_conversion}

\begin{figure}
\centering
\includegraphics[width=0.45\textwidth]{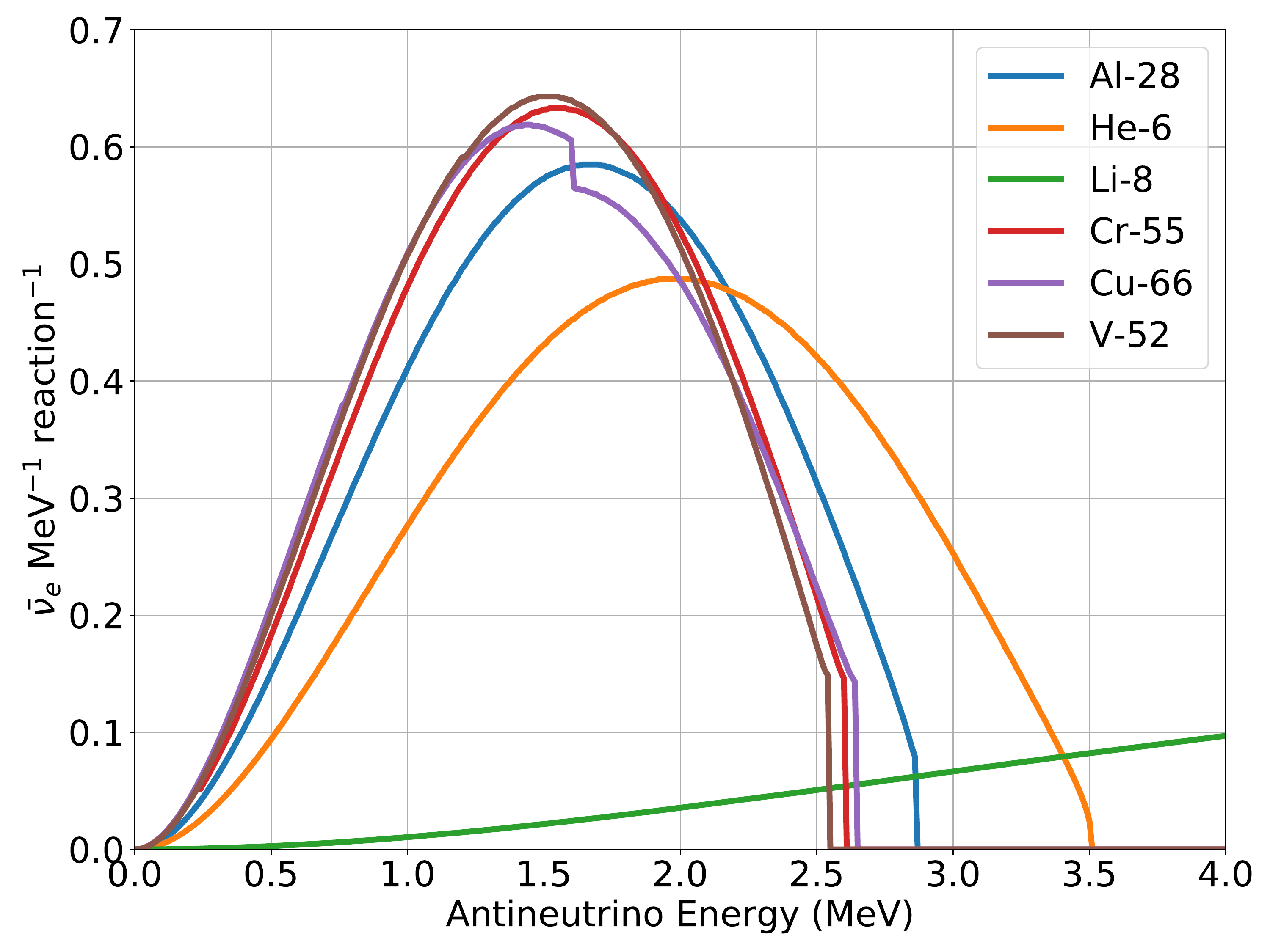}
\caption{Probability density function of \nuebar{} for nonfission candidates. Note that \cu{} has an additional 9\%  $\beta^-$ branch not mentioned in Table~\ref{tab:antinu_candidates}, and the \li{} endpoint is approximately 13~MeV.}
\label{fig:cadidates_nu_spectra}
\end{figure}

The goal of this section is to take the reaction rates calculated from the previous section and convert to \nuebar{} spectra for candidates of interest. The Oklo nuclide tool kit \cite{Oklo} is used to generate \nuebar{} for \uFive{} and the candidates. Oklo uses transition and energy level data from ENSDF-6 \cite{ENSDF} and cumulative fission yield data from the and Evaluated Nuclear Data File (ENDF) \cite{ENDF7}. Both of these are combined to calculate \nuebar{} spectra from fissile isotopes. The Oklo calculation for \nuebar{} spectra includes terms and corrections from several sources \cite{Wilkinson1989,Huber2011,Schenter1983}. The ENSDF data alone can be used to generate \nuebar{} spectra for individual $\beta^-$ decays. Summation predictions of \nuebar{} spectra, such as those produced by Oklo, can have uncertainties as high as 10\% \cite{Huber2016}.

The most commonly used reference reactor \nuebar{} spectrum is that generated by Huber via conversion of experimentally measured a reactor electron spectrum \cite{Huber2011}. The \uFive{} spectra generated by Oklo from Huber have small differences. The most notable differences are in the lower energies (below the IBD threshold) and therefore not of primary interest for this work. The theoretical predictions from Oklo return the fission \nuebar{} spectra in 10~keV bins.

The end product is a prediction of the excess \nuebar{} that are produced from candidates with respect to those from fuel fissions. The excess \nuebar{} from candidates is calculated by taking the ratio of reaction rate of candidate \textit{X} to the \uFive{} fission rate and multiplying by the ratio of \nuebar{} produced per reaction above the IBD threshold ($N_{\nuebar{}}$):

\begin{equation}
\frac{\bar{\nu}\textsubscript{cand}(E)}{\bar{\nu}\textsuperscript{fuel}(E)} = \frac{^A_ZX\textnormal{(n,capture)}}{^{235}\textnormal{U(n,fission)}} \frac{N_{\bar{\nu},X}(E)}{N_{\bar{\nu},\textsuperscript{235}\textnormal{U}}(E)}
\label{eqn:excess_antinu}.
\end{equation}

\noindent In this equation, $N_{\nubar}$ is the number of \nuebar{} produced above the IBD threshold per reaction. Because the fission rate is the most frequent neutron-induced transmutation in a reactor and the fact that fission always produces more \nuebar{} than a single $\beta^-$ decay, both ratios will always be less than unity. The result will be a fraction, or excess, of \nuebar{} above threshold produced by the candidate versus those from the fission process.

Figure \ref{fig:cadidates_nu_spectra} shows the \nuebar{} spectra for the nonfissile candidates (i.e., not including NpO$_2$ and  curium oxide [CmO] targets) from a single $\beta^-$ decay, i.e., the spectra $N_{\bar{\nu},X}(E)$ in Equation~\ref{eqn:excess_antinu}. Lithium-8 is the only candidate with a \nuebar{} endpoint above 3.5~MeV. The \cu{} distribution experiences a dip because there is a 9\% branch that ends at 1.6 MeV. Most distributions have an average \nuebar{} energy lower than the IBD threshold. The two exceptions are \he{} and \li{}, the two products produced in the beryllium reflector.

The next several sections discuss each of the relevant candidates in detail, quantify their decay rates in the reactor as a function of time, and calculate the antineutrino spectrum. Some candidates will then be eliminated from consideration. The elements are grouped into three sections according to purpose listed in Table~\ref{tab:antinu_candidates}: structural (Section~\ref{sec:structural}), reflector (Section~\ref{sec:be}), and targets (Section~\ref{sec:targets}). Reaction rates and activites are calculated in these sections. The conversion to \nuebar{} spectrum and contributions of these isotopes relative to the fission spectrum is discussed in Section~\ref{sec:avg_nonfuel_contributions}.

\subsection{Structural}
\label{sec:structural}

The most prominent structural materials in HFIR include Al, Cu, Cr, and Mn. Aluminum is included in the form of Al-6061, Al-1100, and several others. When HFIR was designed, aluminum was selected because of its low fabrication and reprocessing costs \cite{Cheverton}. It also has a lower reactivity penalty than other structural materials; the only exception is zirconium, which is typically more expensive but more often used in commercial reactors as cladding. Copper, chromium, and manganese are present in much lower quantities in the core than aluminum.

\subsubsection{Aluminum}

Aluminum is the most prominent structural material in HFIR. The natural abundance of aluminum is 100\% $^{27}$Al. In the FTT region, aluminum makes up dummy targets, target rod rabbit holders in the target positions, and capsule bodies. In the IFE and OFE, it is the largest atomic contributor in the U$_3$O$_8$-Al fuel and constitutes most of the filler material, which is the nonfuelled region located within the aluminum cladding \cite{Ilas2015}. The unfueled regions of the fuel plates and side walls of the IFE/OFE are also predominately composed of aluminum. It exists in all regions of the control elements, although absorption is dominated by neutron poisons. Some of the reflector support and HB tube cells are also of relevance.

The reaction of interest for aluminum is $^{27}$Al(n,$\gamma$)\al{} with a $\beta^-$ transition to $^{28}$Si \cite{Al28}. The transition releases 4.642~MeV and results in an excited state of $^{28}$Si at 1.779~MeV; therefore the $\beta^-$ endpoint energy is 2.864~MeV. The half-life of \al{} is 2.245~minutes; therefore, it is assumed the \al{} activity reaches equilibrium quickly into the cycle.

\begin{figure}
\centering
\includegraphics[width=0.45\textwidth]{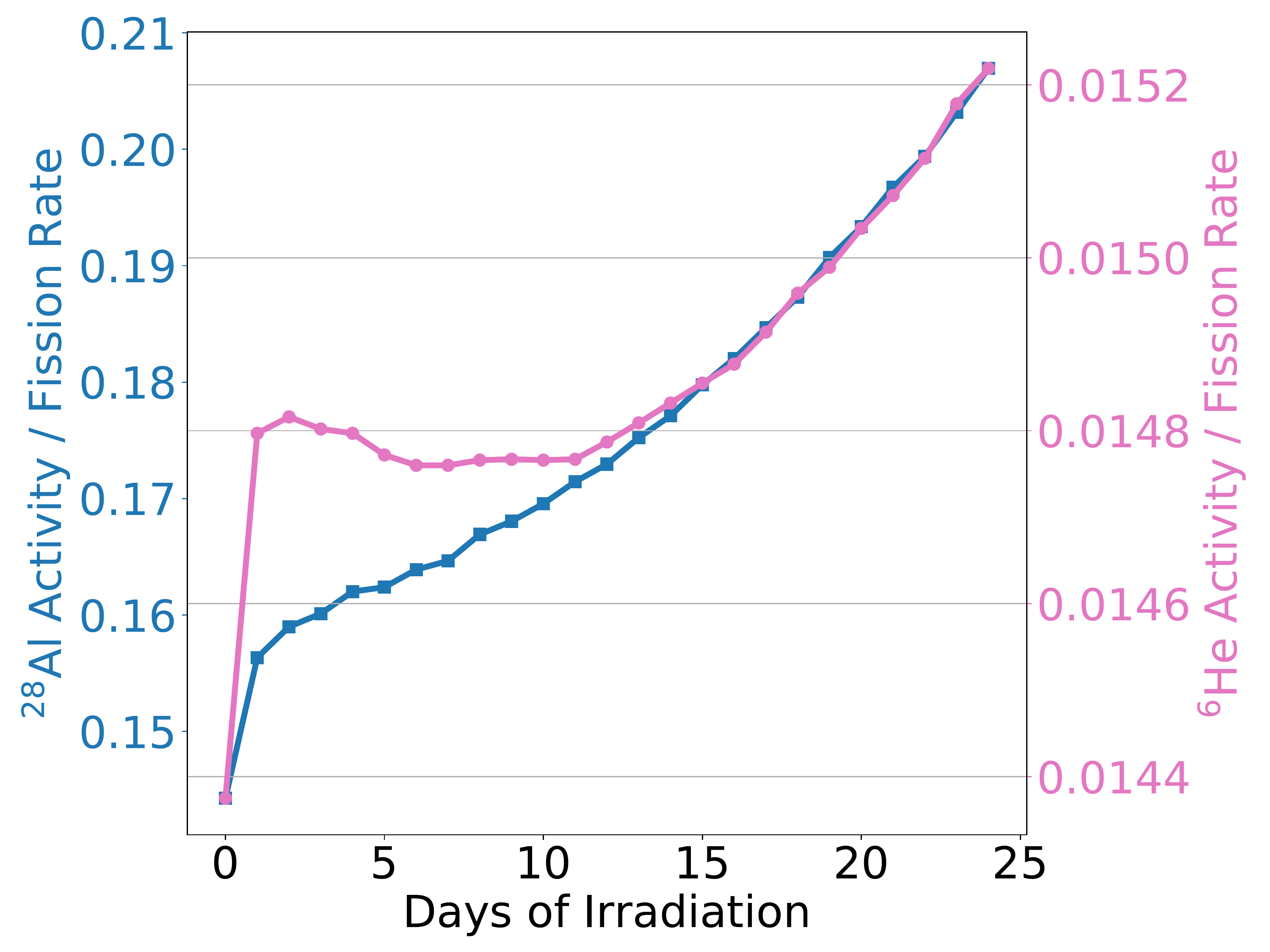}
\caption{\al{} and \he{} activities to fuel fission rate ratio for each day in the cycle}
\label{fig:hfir_al_he}
\end{figure}

In the explicit representative HFIR MCNP model, aluminum is contained in 1,967 cells and the mass is approximately 250~kg. The $^{27}$Al(n,$\gamma$) core activity is calculated according to Equation~\ref{eqn:reaction_rate} and ranges from 4.0 to 5.4 $\times 10^{17}$ s$^{-1}$ from BOC to EOC. These values equate to approximately 15--20\% of the fuel fission rate, as shown in Figure~\ref{fig:hfir_al_he}. The increase throughout the cycle is mostly due to the flux increase in many regions of the core and withdrawal of the CEs; the shape mirrors the CE withdrawal curves in Ref.~\cite{Chandler2016}. The regions that contribute the most to the \al{} activity include the IFE/OFE sidewalls, structures in the FTT, reflector container, and the white (minimally absorbing) regions of the control elements \cite{Conant2018physor}.

A COUPLE-ORIGEN model of each of the aluminum cells is created to compare to MCNP and to evaluate the depletion of aluminum throughout a cycle. The 44-group neutron flux from MCNP for each cell for each day in the cycle is input into COUPLE-ORIGEN to generate time-dependent activities. There were some differences between the MCNP and COUPLE-ORIGEN models, but the cycle average difference was 2\% between the two models. The choice of neutron energy-group structure had little impact on the \al{} activities because nearly all captures occur in the thermal range. Most cells deplete less than 0.01\% from BOC to EOC. The main exception is fuel structural materials, which deplete in aluminum by more than 1\% per cycle, yet the fuel assemblies are replaced every cycle. 

\subsubsection{Chromium, Copper, and Manganese}

Chromium, copper, and manganese are also structural material candidates. Most of these include the steel of the target rod rabbit holder--bearing capsules, the stainless steel ends, and trace amounts in Al-6061 materials in HB tubes and IFE/OFE sidewalls. For these particular elements, only the EOC reaction rates are calculated in MCNP. Because flux in most core regions is higher at EOC than BOC and because most nonfuel materials are not depleted significantly from BOC to EOC, these calculations are considered to be a conservative overestimate of their average \nuebar{} emissions.

Chromium-55 is produced from the (n,$\gamma$) reaction on $^{54}$Cr, which has the lowest abundance and cross section of the four naturally occurring isotopes. The half-life of \cro{} is 3.497~minutes. The $\beta^-$ transition releases 2.603 MeV. Although \cro{} decays to several excited states of $^{55}$Mn, the most probable ($>99.5\%$) is the ground state \cite{Junde2008}. The $\beta^-$ endpoint energy is thus assumed to be 2.603~MeV. Chromium is contained in 221 cells of the model, totalling 16~g. The EOC \cro{} activity is found to be $1.6 \times 10^{13}$ s$^{-1}$, which is lower than the fission rate by a factor of 10$^5$ and therefore rules out \cro{} as a candidate.

Copper-66 is produced from the (n,$\gamma$) reaction on $^{65}$Cu, which has the lower abundance and cross section of the two naturally occurring isotopes. The half-life of \cu{} is 5.120~minutes. The $\beta^-$ transition releases 2.640 MeV. The only transition to the ground state of $^{66}$Zn that has a $\beta^-$ endpoint energy above the IBD threshold occurs approximately 90.77\% of the time \cite{Browne2010}. Copper is contained in 869 cells of the model, totalling 161~g. The EOC \cu{} activity is $1.13 \times 10^{15}$ s$^{-1}$. This is approximately 0.04\% of the fission rate. This results in an excess of no more than 0.02\% in 10 keV \nuebar{} bins; this value is small enough to rule out \cu{} as a candidate.

Manganese-56 is produced from the (n,$\gamma$) reaction on $^{55}$Mn, which is the sole naturally occurring isotope. The half-life of \mn{} is 2.578 hours. The $\beta^-$ transition releases 3.695 MeV. The main transition of interest from \mn{} to $^{56}$Fe is to the 0.846 MeV excited state, which occurs 56.6\% of the time \cite{Junde2011}. The $\beta^-$ endpoint energy for this transition is therefore 2.849 MeV. Manganese is present in 226 cells of the model, totalling 109~g. The EOC \mn{} activity is $5.16 \times 10^{15}$ s$^{-1}$. Because the endpoint energy is low compared to other candidates and the reaction rate ratio is comparable to that of \cu{}, \mn{} is also ruled out as a candidate.

\subsection{Beryllium Reflector}
\label{sec:be}

The beryllium reflector region is the outermost radial region of the core. A fresh RB, SPB, or PB contains almost exclusively beryllium ($>$99\% atomically). The beryllium builds up reaction  products, including neutron poisons $^3$He and $^6$Li, throughout the many irradiation cycles. The transmutation chain also involves the production of the antineutrino candidates \he{} and \li{}. Owing to the multicycle nature of the poison buildup and the beryllium replacement scheme, MCNP and ORIGEN are both used to generate cycle-dependent isotopics and decay rates from a fresh reflector.

Helium-6 is produced directly from the (n,$\alpha$) reaction on beryllium-9 with a neutron threshold of 0.67~MeV. It is the precursor reaction to the production of both neutron poisons. The half-life of \he{} is 0.806~seconds. The released and $\beta^-$ endpoint energy are both 3.507~MeV because all $^6$He decays to the ground state of $^6$Li \cite{Tilley2003}. The $^9$Be(n,$\alpha$) rate during the cycle in the entire reflector ranges from 3.80 to $4.05 \times 10^{15}$ s$^{-1}$, which is shown in Figure~\ref{fig:hfir_al_he}. The \he{} increase is sharper than that for \al{} because of the higher dependence of neutron flux on the CE position, and this behavior follows the CE withdrawal curves \cite{Chandler2016}. The increase is largely caused by the CE withdrawal because there is a harder neutron spectrum at the axial ends of the reflector which increases the (n,$\alpha$) reaction rate. Helium-6 activity decreases by no more than 1\% between cycles due to the buildup and neutron absorption on $^6$Li; therefore, it is relatively independent of cycle and age of reflector regions.

Unlike the cycle-independent activity of \he{}, the lithium isotopes rely heavily on the number of cycles irradiated. The $^6$Li increases in concentration until it reaches equilibrium after five cycles. Because of the overwhelming (n,$\alpha$) cross section of $^6$Li, the higher isotopes $^7$Li and \li{} increase slowly and linearly with irradiation time from the lower-probability neutron capture. The \li{} activity linearly increases to approximately 10$^{12}$~Bq after 50 cycles which is six orders of magnitude less than the fission rate. The RB, which is the most frequently replaced and has the largest proportion of \he{} activity, is replaced around this cycle limit.

In summary, \he{} does produce significant activity relative to the fission rate. Although the \li{} has a large $\beta^-$ endpoint energy, it pales in comparison to the fission reaction rate by a factor of $10^6$. Thus, the \li{} is not considered as a candidate. Further studies can be performed to quantify intentional production of \li{} from lithium-filled target regions for high-energy \nuebar{} spectrum \cite{Conant2019}.

\subsection{Target Materials}
\label{sec:targets}

The three main target material candidates are vanadium and the two actinide-containing targets recently irradiated in HFIR, CmO and NpO$_2$. Vanadium is a common material irradiated in the flux trap. The two actinide targets are used for isotope production. Table~\ref{tab:target_loadings} shows the loadings of the two types of actinide targets for the four most recent HFIR cycles. The actinide targets are usually irradiated for multiple cycles to produce the isotopes desired.

\begin{table}[ht]
\centering
\caption{Loading of materials in cycles of HFIR for CmO and NpO$_2$ (number of target positions filled) with previous number of cycles irradiated in parentheses and vanadium (total grams in FTT).}
\label{tab:target_loadings}
\begin{tabular}{|c|c|c|c|c|}
\hline
Cycle  & Dates (MM/DD/2018) & CmO (\#)   & NpO$_2$ (\#) & V (g) \\
\hline
 479 & 05/01 to 05/25   & 4 (0) & 9 (0) & 274 \\
 480 & 06/17 to 07/06   & 4 (1) & 9 (1) & 260  \\
 481 & 07/24 to 08/17   & 4 (2) & 0 & 228      \\
 482 & 09/04 to 09/28   & 4 (3) & 9 (2) & 248 \\
\hline
\end{tabular}
\end{table}

\subsubsection{Vanadium}
\label{sec:va}

Vanadium is a target material that is primarily irradiated in the FTT region. The representative model \cite{Chandler2016} contains many vanadium-bearing targets. Many of these targets are not solely composed of vanadium as a target material; the representative model contains many generic homogeneous targets to obtain representative loading of elements. The FTT region also has some vanadium capsules in the PTPs and target rod rabbit holders that make up part of its composition. Since PROSPECT has begun taking data, the loading of vanadium in the FTT region has not changed drastically.

Vanadium-52 is produced from the (n,$\gamma$) reaction on \vaFiftyOne{}, which is the main naturally occurring isotope. The only other naturally occurring isotope is \vaFifty{}, which constitutes 0.25\% of vanadium in nature and is not a candidate. The cross section for neutron capture on \vaFifty{} is approximately an order of magnitude higher than that of \vaFiftyOne{}. Capture tallies in vanadium materials showed that the ratio of captures in \vaFifty{} to \vaFiftyOne{} roughly follows this product of abundance and cross section, i.e., \vaFifty{}(n,$\gamma$)/\vaFiftyOne{}(n,$\gamma$) is approximately 2.5\%. Therefore, assuming natural abundance, most of the neutron captures still occur in \vaFiftyOne{} despite the higher cross section of \vaFifty{}.

The half-life of \va{} is 3.743~minutes. The $\beta^-$ transition releases 3.974~MeV. The main transition is to a 1.434~MeV excited state of $^{52}$Cr, the only transition that has a $\beta^-$ endpoint energy above the IBD threshold, occurs approximately 99.2\% of the time \cite{Dong2015}. The endpoint energy is 2.540~MeV. 

To calculate approximate \nuebar{} rates from \va{}, several simulated loadings of vanadium-bearing generic targets are modeled in several positions in the flux trap; these targets contain vanadium in a similar concentration to that in the V+Ni targets in the representative model \cite{Chandler2016}. Several cases are created at BOC and EOC with full-axial vanadium targets loaded into up to 10 FTT positions. Table~\ref{tab:target_loadings} shows the approximate loading in grams of vanadium (total) in the FTT region for the past four cycles, which has typically been in the range of 200--300~g. The loading in the simulation cases created here have vanadium masses between 150 and 370~g, which covers the entire spread of vanadium loading over the previous five cycles.

The capture rates of \vaFiftyOne{} (and \vaFifty{}) are calculated on a per-gram basis for the various cases at both BOC and EOC. Linear regression is performed for the capture rate of \vaFiftyOne{} as a function of mass in the FTT region for both BOC and EOC with a correlation coefficient $> 0.99$. The number of grams from the four cycles can be used to calculate approximate \va{} activities at BOC and EOC from the linear regression. The rates range from 1.58 to $1.82 \times 10^{16}$ s$^{-1}$ for the minimum loading and from 1.70 to $1.95 \times 10^{16}$ s$^{-1}$ for the maximum loading of the previous four cycles.

\subsubsection{Curium}

Targets made of CmO have been irradiated in the FTT region to produce $^{252}$Cf in many recent cycles. The CmO targets take up the full length of the active fuel region. Although the primary actinide composition in the targets is Cm, they also contain smaller concentrations of Pu and Am \cite{Chandler2016}.

Calculations of CmO fission and heat generation rates have been performed at HFIR for safety analysis. The cycle-dependent fission rates of the CmO targets are obtained and analyzed. The fission rates in the targets are dominated by the fission of $^{245}$Cm and $^{247}$Cm, which account for more than two-thirds of the CmO fission rates. Plutonium-241 and californium-251 each contribute at the 5--12\% level. The fission yield data are not available for $^{247}$Cm in ENDF or other databases.

The representative model contains five CmO targets, all near the center of the flux trap \cite{Chandler2016}. The average fission rates among the five targets is between $5.11 \times 10^{14}$ (BOC) and $3.55 \times 10^{14}$ (EOC) s$^{-1}$. This is roughly 0.01--0.02\% of the total core fission rate. Even with five such targets in the flux trap, which is considered typical for a production campaign, the fraction relative to the \uFive{} fission rate would be approximately 0.1\%. The isotopes that contribute most to this fission rate are $^{245}$Cm and $^{247}$Cm. It is assumed that the change to the \uFive{} spectrum would be relatively unaffected by curium fissions. The fission yield differences for the rest of the known isotopes is not significant enough to consider the curium target isotopes as candidates. Note, these targets were analyzed for one cycle but are typically irradiated for many. The total target fission rates decrease with each subsequent cycle so this is deemed to be a conservative estimate of multicycle CmO target irradiations.

\subsubsection{Neptunium}

Neptunium oxide (NpO$_2$) targets have been irradiated in several past cycles to produce \pEight{} for NASA. The targets are irradiated in the VXFs for nominally three cycles. The fission rates in the NpO$_2$ targets are dominated by two isotopes: \pNine{} and \npEight{}. The \npEight{} dominates for the first two cycles, and \pNine{} becomes the dominant contributor at the beginning of the third cycle.

The PROSPECT experiment collected data during three NpO$_2$ irradiation cycles. Nine VXFs were filled with NpO$_2$ targets starting in Cycle~479 and continued into Cycle~480. Cycle~481 contained zero targets with Np and Pu. Cycle 482 continued with the targets' third and final irradiation cycle to date, which is shown in Table~\ref{tab:target_loadings}.

\begin{figure}
\centering
\includegraphics[width=0.45\textwidth]{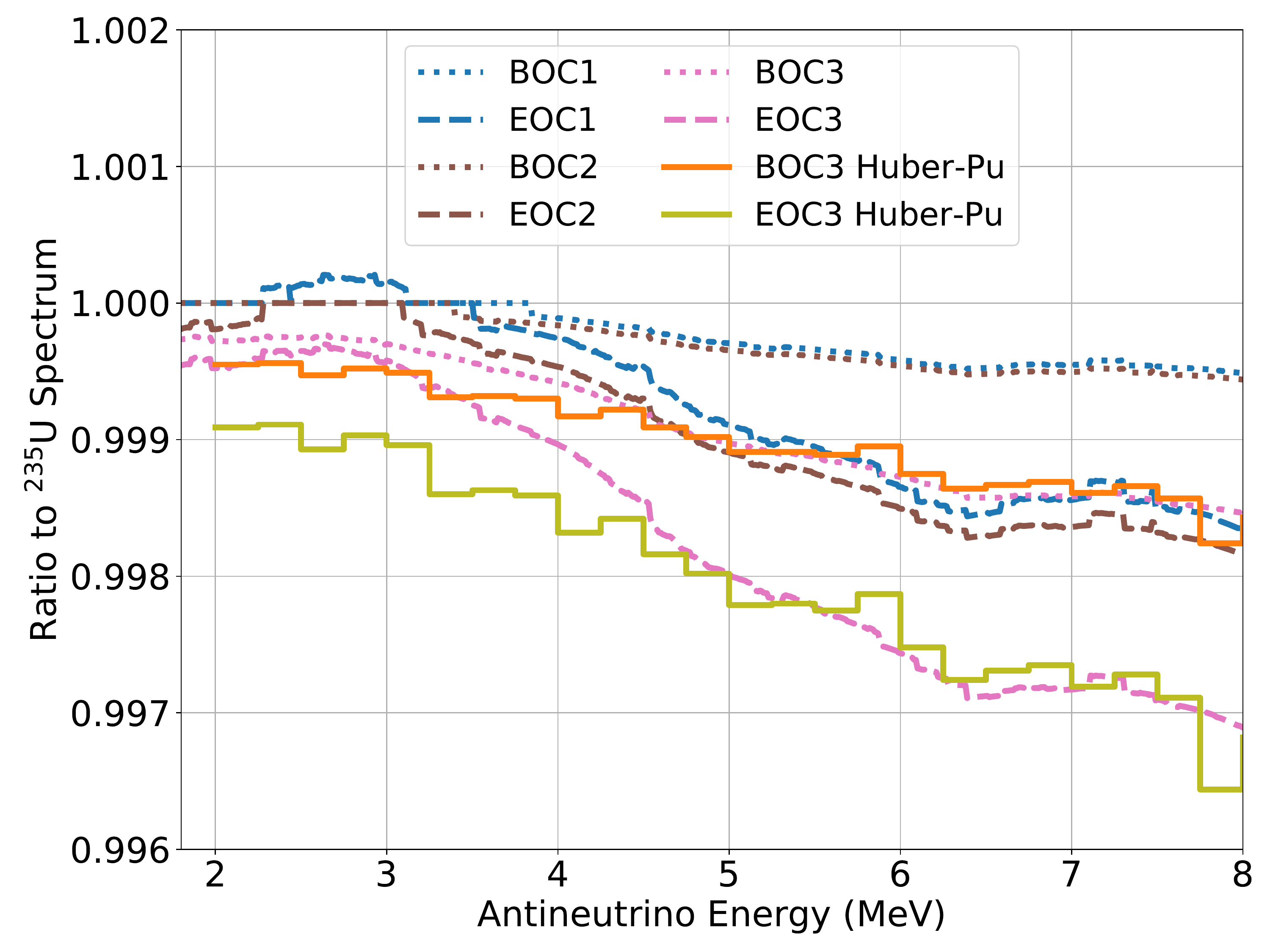}
\caption{\nuebar{} spectrum changes from \uFive{} based on Oklo for beginning and end of a typical three-cycle irradiation of the NpO$_2$ targets using the summation method. BOC/EOC3 Huber-Pu includes differences in the third irradiation cycle between the inclusion of \pNine{} only (no \npEight{}) using Huber predictions.}
\label{fig:np_pu_contribution}
\end{figure}

The Np and Pu fission rates are converted to \nuebar{} spectra using the ENSDF and fission yield data and compared with the \uFive{} nominal spectrum of HFIR. The \npEight{} \nuebar{} spectrum was calculated using Oklo, and its resulting spectrum is comparable to that of \uFive{} but higher by 4--8\% in the 2--6~MeV energy range. The reaction rate ratio of target to fuel fission rate is converted to relative \nuebar{} production rate in a way that is similar to that used in Equation~\ref{eqn:excess_antinu}. Heat power in the reactor is maintained at 85~MW by decreasing the fission rate of \uFive{} to offset the target (Np and Pu) fission rate; this is assumed to be valid because the fission energy release is comparable for the actinides. Note, the core power at HFIR has uncertainties of 2\% due to instrument uncertainty \cite{Conant2019}. Figure~\ref{fig:np_pu_contribution} shows the relative change to the nominal \uFive{} \nuebar{} spectrum for the three cycles of irradiation at BOC/EOC.

To only examine the impact of widely used \pNine{} spectra, only the \pNine{} fission rates in the targets are compared to that for \uFive{} in the fuel using the Huber--Mueller data. The ratio of \pNine{} fissions is highest in the their third cycle of irradiation, so this case is considered for the maximum difference from the nominal \uFive{} spectrum. With the inclusion of the nine NpO$_2$ VXFs, each containing seven targets in their third cycle of irradiation, when the \pNine{} contribution is the highest, the \nuebar{} spectrum decreases by no more than 0.35\% in any energy bin according to the Huber data. This difference is shown in Figure~\ref{fig:np_pu_contribution}. The decrease in the spectrum below the bump region is largely a result of the fission rate and lower \nuebar{} yield of \pNine{}.

\subsection{Cycle Average Nonfuel Contribution to \nuebar{} Spectrum}
\label{sec:avg_nonfuel_contributions}

The results presented so far show that \al{}, \he{}, and \va{} are the most significant candidates of nonfissile \nuebar{} in HFIR. The ratio of \nuebar{} spectrum, according to Equation \ref{eqn:excess_antinu}, is used to calculate cycle-average excess from the selected \nuebar{} candidates. For aluminum and helium, the cycle-average reaction rate is used. For vanadium, an activity corresponding to an average loading in the flux trap is used.

\begin{figure}
\centering
\includegraphics[width=0.45\textwidth]{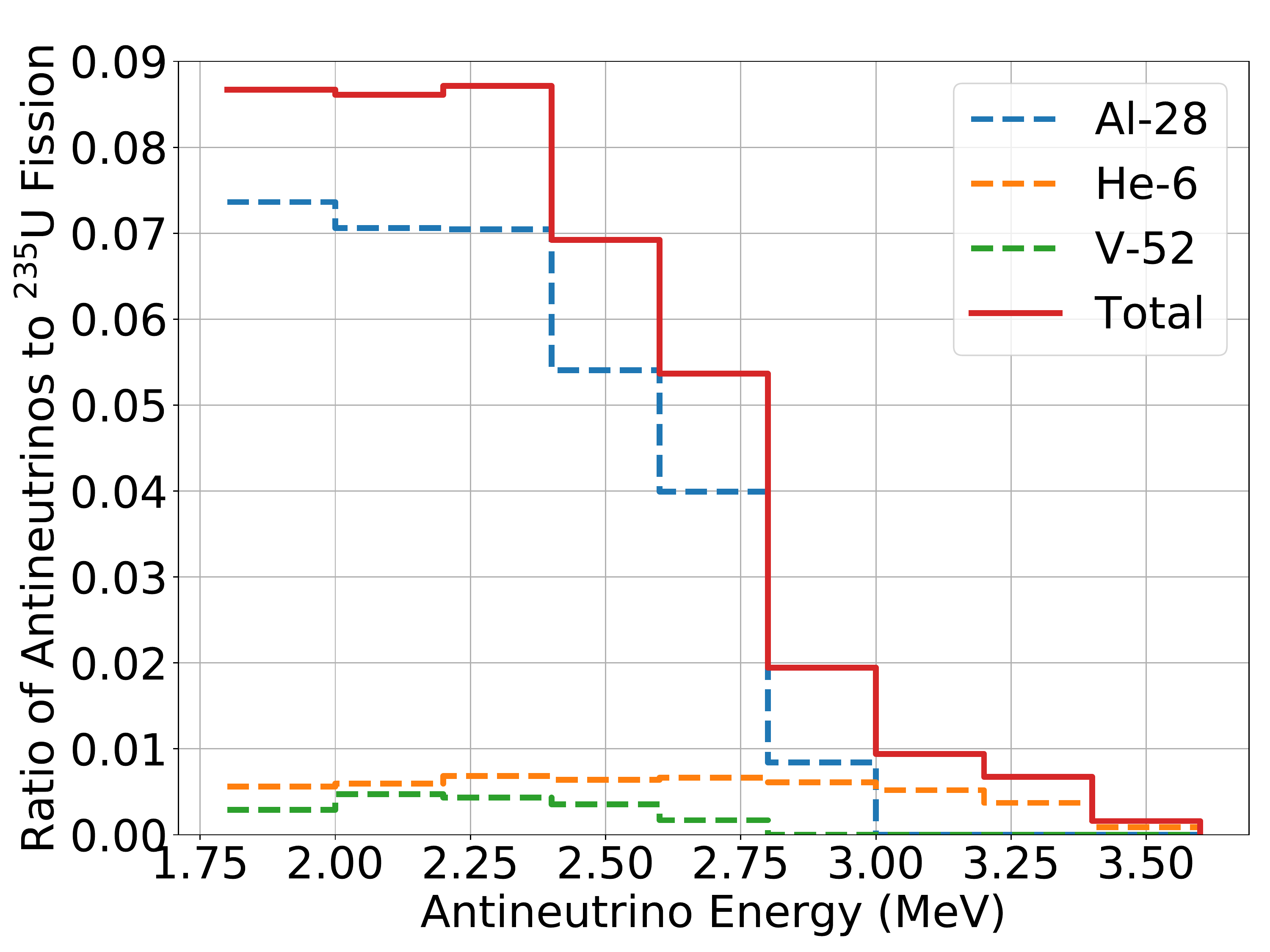}
\caption{ Average excess of \al{}, \he{}, and \va{} contributions to the \nuebar{} spectrum.}
\label{fig:al_he_v}
\end{figure}

Figure \ref{fig:al_he_v} shows the excess contributions in 200~keV bins for the three largest contributions. Aluminum-28 contributes over 8\% in the low-energy range and all three isotopes combine to more than 9\%. The \al{} had by far the largest contribution between 1.8 and 2.86~MeV, its \B{} endpoint. The \he{} has a peak contribution of 0.5--0.75\% effect around 2.5~MeV but drops toward its endpoint 3.5~MeV. The \va{} contribution peaks at about 0.5\%, and its endpoint is comparable to \al{}. In total, these three isotopes increase the expected magnitude of detected reactor spectra by 1\%.

\section{Note on Commercial Reactor Comparisons}
\label{sec:lwr_comparison}

Most reactor \nuebar{} measurements have been collected at commercial nuclear power plants, mainly light water reactors (LWRs). The natural question arises of how nonfuel \nuebar{} may affect the spectrum for a commercial LWR compared to HFIR. A full analysis was not performed, but some insight can be provided based on this analysis. The larger core size and lack of significant experimental facilities at commercial reactors results in less neutron activation of nonfuel materials on a per-fission basis. Commercial LWRs also have a small variety of materials that are contained in the core. The primary nonfuel materials that exist in commercial LWRs include Zircaloy as a cladding material and variations of stainless steels in support structures such as the reactor pressure vessel.

Almost all of the main LWR isotopes of iron and zirconium would be ruled out by the \nuebar{} candidate selection process (Section~\ref{sec:selection}); the only exception is $^{96}$Zr, the isotope of zirconium with the lowest natural abundance. The $^{96}$Zr(n,$\gamma$)$^{97}$Zr transition has only one, albeit dominant, transition that results in a $\beta^-$ endpoint (1.915~MeV) slightly higher than the IBD threshold \cite{Nica2015}. This transition has a half-life of 16.749~hours, which is not negligible but longer that that of most isotopes considered in this work.

Chromium has one neutron capture reaction that results in a \nuebar{} above the IBD threshold, $^{55}$Cr. Its precursor, $^{54}$Cr, is the isotope with the lowest abundance and cross section of chromium isotopes, shown in Table~\ref{tab:antinu_candidates}. Chromium can be contained in 300 series stainless steels, most commonly 304, 308, and 309 in the core structural and pressure vessel \cite{Zinkle2009}. These forms of steel can have between 15\% and 20\% chromium by mass \cite{PNNLcompendium}. Case studies for individual reactors and their chromium content can be performed should precise \nuebar{} predictions be needed.

In summary, \nuebar{} contributions from the minor isotopes of zirconium and chromium in LWRs are estimated to be at least three orders of magnitude lower than that of aluminum in HFIR. Further studies can be done to examine the activation of zirconium or other isotopes (e.g., the chromium composition in steels for specific commercial reactors). This effect is estimated to be small due to the lower ratio of absorption rate to fission rate and the lack of large quantities of chromium in the higher flux regions of the core (i.e., near the center). The nonfuel contributions to the \nuebar{} spectrum should not be a cause for concern for experiments at commercial reactors, such as Daya Bay, Double Chooz, and RENO.

\section{Conclusions}
\label{sec:conclusions}

HFIR's missions allow for a wide variety of different materials to be deliberately or indirectly transmuted to $\beta^-$ decaying products during operation. Potential candidates are examined to find the largest emitters of \nuebar{} that need to be accounted for in the \uFive{} spectrum from HFIR.

A methodology was created to select \nuebar{} candidates from nonfuel materials in HFIR that would contribute nominally to the \nuebar{} spectrum. Several candidates are identified as potentially problematic for the \nuebar{} measurement based on their abundance in the core, cross section, and $\beta^-$ endpoint energy. Reactor simulations were  performed to calculate reaction rates and \nuebar{} spectra from the nonfuel materials. 

The most dominant nonfuel contributors to the \nuebar{} spectrum are the \al{} from structural materials and \he{} from interactions in the beryllium reflector. Both of these \nuebar{} contributions were found to be relatively cycle independent and to increase with cycle time because of the flux increase in many regions of the reactor. The contribution to the \nuebar{} energy spectrum was calculated. Averaged over a cycle, the \al{} dominates with a maximum 7\% contribution near threshold to about 1\% at its \B{} endpoint. The \he{} has a nearly uniform 0.5--0.75\% contribution up until its endpoint. Based on typical loadings in the flux trap, the \va{} has a 0.25--0.5\% contribution. For all energy ranges, these contributions combine for 1\% effect in the total detected \nuebar{}. Such contributions should be calculated for reactors with comparable amounts of aluminum or similar reflector design to support future neutrino experiments.

The contributions of target materials have a high dependence on the amount and location of loading in the core. Vanadium is identified as the target material that was calculated to have as high as a 0.26--0.51\% in the low-energy \nuebar{} range. The irradiation of NpO$_2$ targets has a small but non-negligible impact on the \nuebar{} spectrum. The effect of the recent loading of nine VXF positions with multicycle irradiations of NpO$_2$ yield a maximum of 0.35\% relative change to the nominal \uFive{} spectrum at high energy. Should HFIR irradiate more targets or irradiate them more than three cycles, it would be necessary to analyze further the contribution of \npEight{} and \pNine{} because the \uFive{} fuel fission rate will decrease as a result of heat power conservation. The multicycle NpO$_2$ targets contribution to the spectrum would be exacerbated with subsequent cycles irradiated because of the increase in \pNine{} fission rate and its low \nuebar{} yield compared to \uFive{}. The CmO targets generally would not contribute significantly unless large discrepancies between Cm or Cf and \uFive{} \nuebar{} spectra were discovered.

In summary, this analysis shows that nonfuel reactions make significant contributions to the \nuebar{} spectrum at HFIR. In particular, \al{}, \he{}, and \va{} contributions should be included in the analysis for a PROSPECT-like experiment at HFIR. We suggest that reactor modeling for research reactors may be necessary in the development and analysis of short-baseline antineutrino experiments to account for variations in research reactor design. Although we only examined HFIR in detail, other nonfuel emission candidates may need to be considered depending on reactor composition and missions. 

The findings for these isotopes in HFIR are factored into the PROSPECT detector response matrix. Integrated over the whole \nuebar{} spectrum, the contributions of \al{} and \he{} combined are found to have 1\% effect on the total \nuebar{} flux \cite{PROSPECTprl2019}. For HFIR specifically, nonfuel contributions are not in the energy range high enough to contribute to the bump in the measured spectra. \\

\section{Acronyms}
\label{sec:acronyms}

\noindent BOC \hfill beginning of cycle \\
CE \hfill control element \\
ENDF \hfill Evaluated Nuclear Data File \\
ENSDF \hfill Evaluated Nuclear Structure Data File \\
EOC \hfill end of cycle \\
FTT \hfill flux trap target region \\
HFIR \hfill High Flux Isotope Reactor \\
IBD \hfill inverse beta decay \\
IFE/OFE \hfill inner/outer fuel element \\
ILL \hfill Institut Laue-Langevin \\
MCNP \hfill Monte Carlo N-Particle \\
ORNL \hfill Oak Ridge National Laboratory \\
ORIGEN \hfill Oak Ridge Isotope Generation \\
PB \hfill permanent beryllium \\
PNF \hfill power normalization factor \\
PROSPECT \hfill Precision Reactor Oscillation and Spectrum \\
RB \hfill removable beryllium \\
SPB \hfill semi-permanent beryllium \\
VXF \hfill vertical experiment facility \\

\section{Acknowledgments}

David Chandler at HFIR was instrumental in helping with reactor modeling efforts. Dan Dwyer is acknowledged for his help using the Oklo code.  Additionally, discussions with Greg Hirtz at HFIR were useful in identifying candidates. The fission rates in the CmO targets were provided by Susan Hogle from ORNL.

This material is based upon work supported by the following sources: US Department of Energy (DOE) Office of Science, Office of High Energy Physics under Award No. DE-SC0016357 and DE-SC0017660 to Yale University, under Award No. DE-SC0017815 to Drexel University, under Award No. DE-SC0008347 to Illinois Institute of Technology, under Award No. DE-SC0016060 to Temple University, under Contract No. DE-SC0012704 to Brookhaven National Laboratory, and under Work Proposal Number SCW1504 to Lawrence Livermore National Laboratory. This work was performed under the auspices of the U.S. Department of Energy by Lawrence Livermore National Laboratory under Contract DE-AC52-07NA27344 and by Oak Ridge National Laboratory under Contract DE-AC05-00OR22725. Additional funding for the experiment was provided by the Heising-Simons Foundation under Award No. \#2016-117 to Yale University.

J.G. is supported through the NSF Graduate Research Fellowship Program and A.C. performed work under appointment to the Nuclear Nonproliferation International Safeguards Fellowship Program sponsored by the National Nuclear Security Administration's Office of International Nuclear Safeguards (NA-241). This work was also supported by the Canada First Research Excellence Fund (CFREF), and the Natural Sciences and Engineering Research Council of Canada (NSERC) Discovery program under grant \#RGPIN-418579, and Province of Ontario.

We further acknowledge support from Yale University, the Illinois Institute of Technology, Temple University, Brookhaven National Laboratory, the Lawrence Livermore National Laboratory LDRD program, the National Institute of Standards and Technology, and Oak Ridge National Laboratory. We gratefully acknowledge the support and hospitality of the High Flux Isotope Reactor and Oak Ridge National Laboratory, managed by UT-Battelle for the U.S. Department of Energy.


\bibliographystyle{apsrev4-1}
\bibliography{main}{}


\end{document}

%% file: pfile_author_feb2020.tex
\author{
A.B.~Balantekin$^{n}$,
H.R.~Band$^{o}$,
C.D.~Bass$^{g}$,
D.E.~Bergeron$^{h}$,
D.~Berish$^{k}$,
N.S.~Bowden$^{f}$,
J.P.~Brodsky$^{f}$,
C.D.~Bryan$^{i}$,
T.~Classen$^{f}$,
A.J.~Conant$^{\text{*},c,i}$,
G.~Deichert$^{i}$,
M.V.~Diwan$^{a}$,
M.J.~Dolinski$^{b}$,
A.~Erickson$^{c}$,
B.T.~Foust$^{o}$,
J.K.~Gaison$^{o}$,
A.~Galindo-Uribarri$^{j,l}$,
C.E.~Gilbert$^{j,l}$,
B.T.~Hackett$^{j,l}$,
S.~Hans$^{a}$,
A.B.~Hansell$^{k}$,
K.M.~Heeger$^{o}$,
B.~Heffron$^{j,l}$,
D.E.~Jaffe$^{a}$,
X.~Ji$^{a}$,
D.C.~Jones$^{k}$,
O.~Kyzylova$^{b}$,
C.E.~Lane$^{b}$,
T.J.~Langford$^{o}$,
J.~LaRosa$^{h}$,
B.R.~Littlejohn$^{e}$,
X.~Lu$^{j,l}$,
J.~Maricic$^{d}$
M.P.~Mendenhall$^{f}$,
R.~Milincic$^{d}$,
I.~Mitchell$^{d}$,
P.E.~Mueller$^{j}$,
H.P.~Mumm$^{h}$,
J.~Napolitano$^{k}$,
R.~Neilson$^{b}$,
J.A.~Nikkel$^{o}$,
D.~Norcini$^{o}$,
S.~Nour$^{h}$,
J.L.~Palomino-Gallo$^{e}$,
D.A.~Pushin$^{m}$,
X.~Qian$^{a}$,
E.~Romero-Romero$^{i,k}$,
R.~Rosero$^{a}$,
P.T.~Surukuchi$^{o}$,
M.A.~Tyra$^{h}$,
R.L.~Varner$^{j}$,
C.~White$^{e}$,
J.~Wilhelmi$^{k}$,
A.~Woolverton$^{m}$,
M.~Yeh$^{a}$,
A.~Zhang$^{a}$,
C.~Zhang$^{a}$,
X.~Zhang$^{f}$
}
\address{$^{a}$Brookhaven National Laboratory, Upton, NY 11973, USA}
\address{$^{b}$Department of Physics, Drexel University, Philadelphia, PA 19104, USA}
\address{$^{c}$George W.~Woodruff School of Mechanical Engineering, Georgia Institute of Technology, Atlanta, GA 30332, USA}
\address{$^{d}$Department of Physics \& Astronomy, University of Hawaii, Honolulu, HA 96822, USA}
\address{$^{e}$Department of Physics, Illinois Institute of Technology, Chicago, IL 60616, USA}
\address{$^{f}$Nuclear and Chemical Sciences Division, Lawrence Livermore National Laboratory, Livermore, CA 94550, USA}
\address{$^{g}$Department of Physics, Le Moyne College, Syracuse, NY 13214, USA}
\address{$^{h}$National Institute of Standards and Technology, Gaithersburg, MD 20899, USA}
\address{$^{i}$High Flux Isotope Reactor, Oak Ridge National Laboratory, Oak Ridge, TN 37830, USA}
\address{$^{j}$Physics Division, Oak Ridge National Laboratory, Oak Ridge, TN 37830, USA}
\address{$^{k}$Department of Physics, Temple University, Philadelphia, PA 19122, USA}
\address{$^{l}$Department of Physics and Astronomy, University of Tennessee, Knoxville, TN 37996, USA}
\address{$^{m}$Institute for Quantum Computing and Department of Physics and Astronomy, University of Waterloo, Waterloo, ON N2L 3G1, Canada}
\address{$^{n}$Department of Physics, University of Wisconsin, Madison, Madison, WI 53706, USA}
\address{$^{o}$Wright Laboratory, Department of Physics, Yale University, New Haven, CT 06520, USA}

%% file: main.bbl
\begin{thebibliography}{72}%
\makeatletter
\providecommand \@ifxundefined [1]{%
 \@ifx{#1\undefined}
}%
\providecommand \@ifnum [1]{%
 \ifnum #1\expandafter \@firstoftwo
 \else \expandafter \@secondoftwo
 \fi
}%
\providecommand \@ifx [1]{%
 \ifx #1\expandafter \@firstoftwo
 \else \expandafter \@secondoftwo
 \fi
}%
\providecommand \natexlab [1]{#1}%
\providecommand \enquote  [1]{``#1''}%
\providecommand \bibnamefont  [1]{#1}%
\providecommand \bibfnamefont [1]{#1}%
\providecommand \citenamefont [1]{#1}%
\providecommand \href@noop [0]{\@secondoftwo}%
\providecommand \href [0]{\begingroup \@sanitize@url \@href}%
\providecommand \@href[1]{\@@startlink{#1}\@@href}%
\providecommand \@@href[1]{\endgroup#1\@@endlink}%
\providecommand \@sanitize@url [0]{\catcode `\\12\catcode `\$12\catcode
  `\&12\catcode `\#12\catcode `\^12\catcode `\_12\catcode `\%12\relax}%
\providecommand \@@startlink[1]{}%
\providecommand \@@endlink[0]{}%
\providecommand \url  [0]{\begingroup\@sanitize@url \@url }%
\providecommand \@url [1]{\endgroup\@href {#1}{\urlprefix }}%
\providecommand \urlprefix  [0]{URL }%
\providecommand \Eprint [0]{\href }%
\providecommand \doibase [0]{http://dx.doi.org/}%
\providecommand \selectlanguage [0]{\@gobble}%
\providecommand \bibinfo  [0]{\@secondoftwo}%
\providecommand \bibfield  [0]{\@secondoftwo}%
\providecommand \translation [1]{[#1]}%
\providecommand \BibitemOpen [0]{}%
\providecommand \bibitemStop [0]{}%
\providecommand \bibitemNoStop [0]{.\EOS\space}%
\providecommand \EOS [0]{\spacefactor3000\relax}%
\providecommand \BibitemShut  [1]{\csname bibitem#1\endcsname}%
\let\auto@bib@innerbib\@empty
\bibitem [{\citenamefont {Heeger}\ \emph {et~al.}(2013)\citenamefont {Heeger},
  \citenamefont {Littlejohn}, \citenamefont {Mumm},\ and\ \citenamefont
  {Tobin}}]{Heeger:2012tc}%
  \BibitemOpen
  \bibfield  {author} {\bibinfo {author} {\bibfnamefont {K.~M.}\ \bibnamefont
  {Heeger}}, \bibinfo {author} {\bibfnamefont {B.~R.}\ \bibnamefont
  {Littlejohn}}, \bibinfo {author} {\bibfnamefont {H.~P.}\ \bibnamefont
  {Mumm}}, \ and\ \bibinfo {author} {\bibfnamefont {M.~N.}\ \bibnamefont
  {Tobin}},\ }\href {\doibase 10.1103/PhysRevD.87.073008} {\bibfield  {journal}
  {\bibinfo  {journal} {Phys. Rev.}\ }\textbf {\bibinfo {volume} {D87}},\
  \bibinfo {pages} {073008} (\bibinfo {year} {2013})},\ \Eprint
  {http://arxiv.org/abs/1212.2182} {arXiv:1212.2182 [hep-ex]} \BibitemShut
  {NoStop}%
\bibitem [{\citenamefont {Kim}(2016)}]{Kim2016}%
  \BibitemOpen
  \bibfield  {author} {\bibinfo {author} {\bibfnamefont {Y.}~\bibnamefont
  {Kim}},\ }\href {\doibase 10.1016/j.net.2016.02.001} {\bibfield  {journal}
  {\bibinfo  {journal} {Nucl. Eng. Technol.}\ }\textbf {\bibinfo {volume}
  {48}},\ \bibinfo {pages} {285} (\bibinfo {year} {2016})}\BibitemShut
  {NoStop}%
\bibitem [{\citenamefont {An}\ \emph {et~al.}(2017)\citenamefont {An} \emph
  {et~al.}}]{DayaBayEvolution}%
  \BibitemOpen
  \bibfield  {author} {\bibinfo {author} {\bibfnamefont {F.}~\bibnamefont {An}}
  \emph {et~al.} (\bibinfo {collaboration} {Daya Bay Collaboration}),\ }\href
  {\doibase 10.1103/PhysRevLett.118.251801} {\bibfield  {journal} {\bibinfo
  {journal} {Phys. Rev. Lett.}\ }\textbf {\bibinfo {volume} {118}},\ \bibinfo
  {pages} {251801} (\bibinfo {year} {2017})}\BibitemShut {NoStop}%
\bibitem [{\citenamefont {Klimov}\ \emph {et~al.}(1994)\citenamefont {Klimov},
  \citenamefont {Kopeikin}, \citenamefont {Mika{\'{e}}lyan}, \citenamefont
  {Ozerov},\ and\ \citenamefont {Sinev}}]{Klimov1994}%
  \BibitemOpen
  \bibfield  {author} {\bibinfo {author} {\bibfnamefont {Y.~V.}\ \bibnamefont
  {Klimov}}, \bibinfo {author} {\bibfnamefont {V.~I.}\ \bibnamefont
  {Kopeikin}}, \bibinfo {author} {\bibfnamefont {L.~A.}\ \bibnamefont
  {Mika{\'{e}}lyan}}, \bibinfo {author} {\bibfnamefont {K.~V.}\ \bibnamefont
  {Ozerov}}, \ and\ \bibinfo {author} {\bibfnamefont {V.~V.}\ \bibnamefont
  {Sinev}},\ }\href {\doibase 10.1007/BF02414355} {\bibfield  {journal}
  {\bibinfo  {journal} {At. Energy}\ }\textbf {\bibinfo {volume} {76}},\
  \bibinfo {pages} {123} (\bibinfo {year} {1994})}\BibitemShut {NoStop}%
\bibitem [{\citenamefont {An}\ \emph {et~al.}(2015)\citenamefont {An} \emph
  {et~al.}}]{DayaBay}%
  \BibitemOpen
  \bibfield  {author} {\bibinfo {author} {\bibfnamefont {F.}~\bibnamefont {An}}
  \emph {et~al.},\ }\href {\doibase 10.1103/PhysRevLett.115.111802} {\bibfield
  {journal} {\bibinfo  {journal} {Phys. Rev. Lett.}\ }\textbf {\bibinfo
  {volume} {115}} (\bibinfo {year} {2015}),\ 10.1103/PhysRevLett.115.111802},\
  \Eprint {http://arxiv.org/abs/1505.03456} {arXiv:1505.03456} \BibitemShut
  {NoStop}%
\bibitem [{\citenamefont {Choi}\ \emph {et~al.}(2016)\citenamefont {Choi} \emph
  {et~al.}}]{RENO}%
  \BibitemOpen
  \bibfield  {author} {\bibinfo {author} {\bibfnamefont {J.}~\bibnamefont
  {Choi}} \emph {et~al.},\ }\href {\doibase 10.1103/PhysRevLett.116.211801}
  {\bibfield  {journal} {\bibinfo  {journal} {Phys. Rev. Lett.}\ }\textbf
  {\bibinfo {volume} {116}} (\bibinfo {year} {2016}),\
  10.1103/PhysRevLett.116.211801},\ \Eprint {http://arxiv.org/abs/1511.05849}
  {arXiv:1511.05849} \BibitemShut {NoStop}%
\bibitem [{\citenamefont {{Y. Abe and others}}(2014)}]{DoubleChooz}%
  \BibitemOpen
  \bibfield  {author} {\bibinfo {author} {\bibnamefont {{Y. Abe and others}}},\
  }\href {\doibase 10.1007/JHEP10(2014)086} {\bibfield  {journal} {\bibinfo
  {journal} {J. High Energy Phys.}\ }\textbf {\bibinfo {volume} {2014}},\
  \bibinfo {pages} {86} (\bibinfo {year} {2014})},\ \Eprint
  {http://arxiv.org/abs/1406.7763} {arXiv:1406.7763} \BibitemShut {NoStop}%
\bibitem [{\citenamefont {Mention}\ \emph {et~al.}(2011)\citenamefont
  {Mention}, \citenamefont {Fechner}, \citenamefont {Lasserre}, \citenamefont
  {Mueller}, \citenamefont {Lhuillier}, \citenamefont {Cribier},\ and\
  \citenamefont {Letourneau}}]{Mention2011}%
  \BibitemOpen
  \bibfield  {author} {\bibinfo {author} {\bibfnamefont {G.}~\bibnamefont
  {Mention}}, \bibinfo {author} {\bibfnamefont {M.}~\bibnamefont {Fechner}},
  \bibinfo {author} {\bibfnamefont {T.}~\bibnamefont {Lasserre}}, \bibinfo
  {author} {\bibfnamefont {T.~A.}\ \bibnamefont {Mueller}}, \bibinfo {author}
  {\bibfnamefont {D.}~\bibnamefont {Lhuillier}}, \bibinfo {author}
  {\bibfnamefont {M.}~\bibnamefont {Cribier}}, \ and\ \bibinfo {author}
  {\bibfnamefont {A.}~\bibnamefont {Letourneau}},\ }\href {\doibase
  10.1103/PhysRevD.83.073006} {\bibfield  {journal} {\bibinfo  {journal} {Phys.
  Rev. D - Part. Fields, Gravit. Cosmol.}\ }\textbf {\bibinfo {volume} {83}}
  (\bibinfo {year} {2011}),\ 10.1103/PhysRevD.83.073006},\ \Eprint
  {http://arxiv.org/abs/1101.2755} {arXiv:1101.2755} \BibitemShut {NoStop}%
\bibitem [{\citenamefont {Huber}(2011)}]{Huber2011}%
  \BibitemOpen
  \bibfield  {author} {\bibinfo {author} {\bibfnamefont {P.}~\bibnamefont
  {Huber}},\ }\href@noop {} {\bibfield  {journal} {\bibinfo  {journal} {Phys.
  Rev. C}\ }\textbf {\bibinfo {volume} {84}},\ \bibinfo {pages} {024617}
  (\bibinfo {year} {2011})}\BibitemShut {NoStop}%
\bibitem [{\citenamefont {Fallot}\ \emph {et~al.}(2012)\citenamefont {Fallot}
  \emph {et~al.}}]{Fallot2012}%
  \BibitemOpen
  \bibfield  {author} {\bibinfo {author} {\bibfnamefont {M.}~\bibnamefont
  {Fallot}} \emph {et~al.},\ }\href {\doibase 10.1103/PhysRevLett.109.202504}
  {\bibfield  {journal} {\bibinfo  {journal} {Phys. Rev. Lett.}\ }\textbf
  {\bibinfo {volume} {109}},\ \bibinfo {pages} {202504} (\bibinfo {year}
  {2012})}\BibitemShut {NoStop}%
\bibitem [{\citenamefont {Hayes}\ \emph {et~al.}(2014)\citenamefont {Hayes},
  \citenamefont {Friar}, \citenamefont {Garvey}, \citenamefont {Jungman},\ and\
  \citenamefont {Jonkmans}}]{Hayes2014}%
  \BibitemOpen
  \bibfield  {author} {\bibinfo {author} {\bibfnamefont {A.~C.}\ \bibnamefont
  {Hayes}}, \bibinfo {author} {\bibfnamefont {J.~L.}\ \bibnamefont {Friar}},
  \bibinfo {author} {\bibfnamefont {G.~T.}\ \bibnamefont {Garvey}}, \bibinfo
  {author} {\bibfnamefont {G.}~\bibnamefont {Jungman}}, \ and\ \bibinfo
  {author} {\bibfnamefont {G.}~\bibnamefont {Jonkmans}},\ }\href {\doibase
  10.1103/PhysRevLett.112.202501} {\bibfield  {journal} {\bibinfo  {journal}
  {Phys. Rev. Lett.}\ }\textbf {\bibinfo {volume} {112}},\ \bibinfo {pages}
  {202501} (\bibinfo {year} {2014})}\BibitemShut {NoStop}%
\bibitem [{\citenamefont {Hayes}\ \emph {et~al.}(2015)\citenamefont {Hayes},
  \citenamefont {Friar}, \citenamefont {Garvey}, \citenamefont {Ibeling},
  \citenamefont {Jungman}, \citenamefont {Kawano},\ and\ \citenamefont
  {Mills}}]{Hayes2015}%
  \BibitemOpen
  \bibfield  {author} {\bibinfo {author} {\bibfnamefont {A.~C.}\ \bibnamefont
  {Hayes}}, \bibinfo {author} {\bibfnamefont {J.~L.}\ \bibnamefont {Friar}},
  \bibinfo {author} {\bibfnamefont {G.~T.}\ \bibnamefont {Garvey}}, \bibinfo
  {author} {\bibfnamefont {D.}~\bibnamefont {Ibeling}}, \bibinfo {author}
  {\bibfnamefont {G.}~\bibnamefont {Jungman}}, \bibinfo {author} {\bibfnamefont
  {T.}~\bibnamefont {Kawano}}, \ and\ \bibinfo {author} {\bibfnamefont {R.~W.}\
  \bibnamefont {Mills}},\ }\href {\doibase 10.1103/PhysRevD.92.033015}
  {\bibfield  {journal} {\bibinfo  {journal} {Phys. Rev. D - Part. Fields,
  Gravit. Cosmol.}\ }\textbf {\bibinfo {volume} {92}} (\bibinfo {year}
  {2015}),\ 10.1103/PhysRevD.92.033015},\ \Eprint
  {http://arxiv.org/abs/1506.00583} {arXiv:1506.00583} \BibitemShut {NoStop}%
\bibitem [{\citenamefont {Dwyer}\ and\ \citenamefont
  {Langford}(2015)}]{DwyerLangford2015}%
  \BibitemOpen
  \bibfield  {author} {\bibinfo {author} {\bibfnamefont {D.~A.}\ \bibnamefont
  {Dwyer}}\ and\ \bibinfo {author} {\bibfnamefont {T.~J.}\ \bibnamefont
  {Langford}},\ }\href {\doibase 10.1103/PhysRevLett.114.012502} {\bibfield
  {journal} {\bibinfo  {journal} {Phys. Rev. Lett.}\ }\textbf {\bibinfo
  {volume} {114}},\ \bibinfo {pages} {012502} (\bibinfo {year}
  {2015})}\BibitemShut {NoStop}%
\bibitem [{\citenamefont {Littlejohn}\ \emph {et~al.}(2018)\citenamefont
  {Littlejohn}, \citenamefont {Conant}, \citenamefont {Dwyer}, \citenamefont
  {Erickson}, \citenamefont {Gustafson},\ and\ \citenamefont
  {Hermanek}}]{Littlejohn2018}%
  \BibitemOpen
  \bibfield  {author} {\bibinfo {author} {\bibfnamefont {B.~R.}\ \bibnamefont
  {Littlejohn}}, \bibinfo {author} {\bibfnamefont {A.}~\bibnamefont {Conant}},
  \bibinfo {author} {\bibfnamefont {D.~A.}\ \bibnamefont {Dwyer}}, \bibinfo
  {author} {\bibfnamefont {A.}~\bibnamefont {Erickson}}, \bibinfo {author}
  {\bibfnamefont {I.}~\bibnamefont {Gustafson}}, \ and\ \bibinfo {author}
  {\bibfnamefont {K.}~\bibnamefont {Hermanek}},\ }\href@noop {} {\bibfield
  {journal} {\bibinfo  {journal} {Phys. Rev. D}\ }\textbf {\bibinfo {volume}
  {97}},\ \bibinfo {pages} {073007} (\bibinfo {year} {2018})}\BibitemShut
  {NoStop}%
\bibitem [{\citenamefont {Schreckenbach}\ \emph {et~al.}(1985)\citenamefont
  {Schreckenbach}, \citenamefont {Colvin}, \citenamefont {Gelletly},\ and\
  \citenamefont {{Von Feilitzsch}}}]{Sch1985}%
  \BibitemOpen
  \bibfield  {author} {\bibinfo {author} {\bibfnamefont {K.}~\bibnamefont
  {Schreckenbach}}, \bibinfo {author} {\bibfnamefont {G.}~\bibnamefont
  {Colvin}}, \bibinfo {author} {\bibfnamefont {W.}~\bibnamefont {Gelletly}}, \
  and\ \bibinfo {author} {\bibfnamefont {F.}~\bibnamefont {{Von Feilitzsch}}},\
  }\href {\doibase 10.1016/0370-2693(85)91337-1} {\bibfield  {journal}
  {\bibinfo  {journal} {Phys. Lett. B}\ }\textbf {\bibinfo {volume} {160}},\
  \bibinfo {pages} {325} (\bibinfo {year} {1985})}\BibitemShut {NoStop}%
\bibitem [{\citenamefont {Kopeikin}(2003)}]{Kopeikin2003}%
  \BibitemOpen
  \bibfield  {author} {\bibinfo {author} {\bibfnamefont {V.~I.}\ \bibnamefont
  {Kopeikin}},\ }\href {\doibase 10.1134/1.1563708} {\bibfield  {journal}
  {\bibinfo  {journal} {Phys. of At. Nuclei}\ }\textbf {\bibinfo {volume}
  {66}},\ \bibinfo {pages} {472} (\bibinfo {year} {2003})}\BibitemShut
  {NoStop}%
\bibitem [{\citenamefont {Mueller}\ \emph {et~al.}(2011)\citenamefont
  {Mueller}, \citenamefont {Lhuillier}, \citenamefont {Fallot}, \citenamefont
  {Letourneau}, \citenamefont {Cormon}, \citenamefont {Fechner}, \citenamefont
  {Giot}, \citenamefont {Lasserre}, \citenamefont {Martino}, \citenamefont
  {Mention}, \citenamefont {Porta},\ and\ \citenamefont
  {Yermia}}]{Mueller2011}%
  \BibitemOpen
  \bibfield  {author} {\bibinfo {author} {\bibfnamefont {T.~A.}\ \bibnamefont
  {Mueller}}, \bibinfo {author} {\bibfnamefont {D.}~\bibnamefont {Lhuillier}},
  \bibinfo {author} {\bibfnamefont {M.}~\bibnamefont {Fallot}}, \bibinfo
  {author} {\bibfnamefont {A.}~\bibnamefont {Letourneau}}, \bibinfo {author}
  {\bibfnamefont {S.}~\bibnamefont {Cormon}}, \bibinfo {author} {\bibfnamefont
  {M.}~\bibnamefont {Fechner}}, \bibinfo {author} {\bibfnamefont
  {L.}~\bibnamefont {Giot}}, \bibinfo {author} {\bibfnamefont {T.}~\bibnamefont
  {Lasserre}}, \bibinfo {author} {\bibfnamefont {J.}~\bibnamefont {Martino}},
  \bibinfo {author} {\bibfnamefont {G.}~\bibnamefont {Mention}}, \bibinfo
  {author} {\bibfnamefont {A.}~\bibnamefont {Porta}}, \ and\ \bibinfo {author}
  {\bibfnamefont {F.}~\bibnamefont {Yermia}},\ }\href {\doibase
  10.1103/PhysRevC.83.054615} {\bibfield  {journal} {\bibinfo  {journal} {Phys.
  Rev. C}\ }\textbf {\bibinfo {volume} {83}},\ \bibinfo {pages} {054615}
  (\bibinfo {year} {2011})}\BibitemShut {NoStop}%
\bibitem [{\citenamefont {Huber}\ and\ \citenamefont
  {Jaffke}(2016)}]{HuberJaffke2015}%
  \BibitemOpen
  \bibfield  {author} {\bibinfo {author} {\bibfnamefont {P.}~\bibnamefont
  {Huber}}\ and\ \bibinfo {author} {\bibfnamefont {D.}~\bibnamefont {Jaffke}},\
  }\href {\doibase 10.1103/PhysRevLett.116.122503} {\bibfield  {journal}
  {\bibinfo  {journal} {Phys. Rev. Lett.}\ }\textbf {\bibinfo {volume} {116}},\
  \bibinfo {pages} {122503} (\bibinfo {year} {2016})}\BibitemShut {NoStop}%
\bibitem [{\citenamefont {Zhou}\ \emph {et~al.}(2012)\citenamefont {Zhou},
  \citenamefont {Ruan}, \citenamefont {Nie}, \citenamefont {Zhou},
  \citenamefont {An},\ and\ \citenamefont {Cao}}]{Zhou2012}%
  \BibitemOpen
  \bibfield  {author} {\bibinfo {author} {\bibfnamefont {B.}~\bibnamefont
  {Zhou}}, \bibinfo {author} {\bibfnamefont {X.-C.}\ \bibnamefont {Ruan}},
  \bibinfo {author} {\bibfnamefont {Y.-B.}\ \bibnamefont {Nie}}, \bibinfo
  {author} {\bibfnamefont {Z.-Y.}\ \bibnamefont {Zhou}}, \bibinfo {author}
  {\bibfnamefont {F.-P.}\ \bibnamefont {An}}, \ and\ \bibinfo {author}
  {\bibfnamefont {J.}~\bibnamefont {Cao}},\ }\href {\doibase
  10.1088/1674-1137/36/1/001} {\bibfield  {journal} {\bibinfo  {journal}
  {Chinese Physics C}\ }\textbf {\bibinfo {volume} {36}},\ \bibinfo {pages} {1}
  (\bibinfo {year} {2012})}\BibitemShut {NoStop}%
\bibitem [{\citenamefont {Brdar}\ \emph {et~al.}(2017)\citenamefont {Brdar},
  \citenamefont {Huber},\ and\ \citenamefont {Kopp}}]{Brdar2017}%
  \BibitemOpen
  \bibfield  {author} {\bibinfo {author} {\bibfnamefont {V.}~\bibnamefont
  {Brdar}}, \bibinfo {author} {\bibfnamefont {P.}~\bibnamefont {Huber}}, \ and\
  \bibinfo {author} {\bibfnamefont {J.}~\bibnamefont {Kopp}},\ }\href {\doibase
  10.1103/PhysRevApplied.8.054050} {\bibfield  {journal} {\bibinfo  {journal}
  {Phys. Rev. Applied}\ }\textbf {\bibinfo {volume} {8}},\ \bibinfo {pages}
  {054050} (\bibinfo {year} {2017})}\BibitemShut {NoStop}%
\bibitem [{\citenamefont {Ashenfelter}\ \emph {et~al.}(2018)\citenamefont
  {Ashenfelter} \emph {et~al.}}]{PROSPECTdetector}%
  \BibitemOpen
  \bibfield  {author} {\bibinfo {author} {\bibfnamefont {J.}~\bibnamefont
  {Ashenfelter}} \emph {et~al.},\ }\href
  {http://stacks.iop.org/1748-0221/13/i=06/a=P06023} {\bibfield  {journal}
  {\bibinfo  {journal} {J. Instrum.}\ }\textbf {\bibinfo {volume} {13}},\
  \bibinfo {pages} {P06023} (\bibinfo {year} {2018})}\BibitemShut {NoStop}%
\bibitem [{\citenamefont {Allemandou}\ \emph {et~al.}(2018)\citenamefont
  {Allemandou} \emph {et~al.}}]{STEREO}%
  \BibitemOpen
  \bibfield  {author} {\bibinfo {author} {\bibfnamefont {N.}~\bibnamefont
  {Allemandou}} \emph {et~al.},\ }\href {\doibase
  10.1088/1748-0221/13/07/p07009} {\bibfield  {journal} {\bibinfo  {journal}
  {J. Instrum.}\ }\textbf {\bibinfo {volume} {13}},\ \bibinfo {pages} {P07009}
  (\bibinfo {year} {2018})}\BibitemShut {NoStop}%
\bibitem [{\citenamefont {Abreu}\ \emph {et~al.}(2018)\citenamefont {Abreu}
  \emph {et~al.}}]{SoLid}%
  \BibitemOpen
  \bibfield  {author} {\bibinfo {author} {\bibfnamefont {Y.}~\bibnamefont
  {Abreu}} \emph {et~al.},\ }\href {\doibase 10.1088/1748-0221/13/05/p05005}
  {\bibfield  {journal} {\bibinfo  {journal} {J. Instrum.}\ }\textbf {\bibinfo
  {volume} {13}},\ \bibinfo {pages} {P05005} (\bibinfo {year}
  {2018})}\BibitemShut {NoStop}%
\bibitem [{\citenamefont {Akimov}\ \emph {et~al.}(2017)\citenamefont {Akimov}
  \emph {et~al.}}]{COHERENT}%
  \BibitemOpen
  \bibfield  {author} {\bibinfo {author} {\bibfnamefont {D.}~\bibnamefont
  {Akimov}} \emph {et~al.},\ }\href {\doibase 10.1126/science.aao0990}
  {\bibfield  {journal} {\bibinfo  {journal} {Science}\ }\textbf {\bibinfo
  {volume} {357}},\ \bibinfo {pages} {1123} (\bibinfo {year} {2017})},\ \Eprint
  {http://arxiv.org/abs/https://science.sciencemag.org/content/357/6356/1123.full.pdf}
  {https://science.sciencemag.org/content/357/6356/1123.full.pdf} \BibitemShut
  {NoStop}%
\bibitem [{\citenamefont {Angloher}\ \emph {et~al.}(2019)\citenamefont
  {Angloher} \emph {et~al.}}]{Chooz2019}%
  \BibitemOpen
  \bibfield  {author} {\bibinfo {author} {\bibfnamefont {G.}~\bibnamefont
  {Angloher}} \emph {et~al.},\ }\href {\doibase 10.1140/epjc/s10052-019-7454-4}
  {\bibfield  {journal} {\bibinfo  {journal} {Europ. Phys. Jour. C}\ }\textbf
  {\bibinfo {volume} {79}},\ \bibinfo {pages} {1018} (\bibinfo {year}
  {2019})}\BibitemShut {NoStop}%
\bibitem [{\citenamefont {Bowden}\ \emph {et~al.}(2009)\citenamefont {Bowden},
  \citenamefont {Bernstein}, \citenamefont {Dazeley}, \citenamefont {Svoboda},
  \citenamefont {Misner},\ and\ \citenamefont {Palmer}}]{Bowden2009}%
  \BibitemOpen
  \bibfield  {author} {\bibinfo {author} {\bibfnamefont {N.~S.}\ \bibnamefont
  {Bowden}}, \bibinfo {author} {\bibfnamefont {A.}~\bibnamefont {Bernstein}},
  \bibinfo {author} {\bibfnamefont {S.}~\bibnamefont {Dazeley}}, \bibinfo
  {author} {\bibfnamefont {R.}~\bibnamefont {Svoboda}}, \bibinfo {author}
  {\bibfnamefont {A.}~\bibnamefont {Misner}}, \ and\ \bibinfo {author}
  {\bibfnamefont {T.}~\bibnamefont {Palmer}},\ }\href {\doibase
  10.1063/1.3080251} {\bibfield  {journal} {\bibinfo  {journal} {Jour. of Appl.
  Phys.}\ } (\bibinfo {year} {2009}),\ 10.1063/1.3080251},\ \Eprint
  {http://arxiv.org/abs/0808.0698} {arXiv:0808.0698} \BibitemShut {NoStop}%
\bibitem [{\citenamefont {Chadwick}\ \emph {et~al.}(2011)\citenamefont
  {Chadwick} \emph {et~al.}}]{ENDF7}%
  \BibitemOpen
  \bibfield  {author} {\bibinfo {author} {\bibfnamefont {M.}~\bibnamefont
  {Chadwick}} \emph {et~al.},\ }\href {\doibase
  https://doi.org/10.1016/j.nds.2011.11.002} {\bibfield  {journal} {\bibinfo
  {journal} {Nuclear Data Sheets}\ }\textbf {\bibinfo {volume} {112}},\
  \bibinfo {pages} {2887 } (\bibinfo {year} {2011})},\ \bibinfo {note}
  {{S}pecial Issue on ENDF/B-VII.1 Library}\BibitemShut {NoStop}%
\bibitem [{\citenamefont {Tepel}(1984)}]{ENSDF}%
  \BibitemOpen
  \bibfield  {author} {\bibinfo {author} {\bibfnamefont {J.}~\bibnamefont
  {Tepel}},\ }\href {\doibase https://doi.org/10.1016/0010-4655(84)90115-2}
  {\bibfield  {journal} {\bibinfo  {journal} {Comput. Phys. Commun.}\ }\textbf
  {\bibinfo {volume} {33}},\ \bibinfo {pages} {129 } (\bibinfo {year}
  {1984})}\BibitemShut {NoStop}%
\bibitem [{\citenamefont {Deniz}\ \emph {et~al.}(2010)\citenamefont {Deniz}
  \emph {et~al.}}]{Deniz2010}%
  \BibitemOpen
  \bibfield  {author} {\bibinfo {author} {\bibfnamefont {M.}~\bibnamefont
  {Deniz}} \emph {et~al.} (\bibinfo {collaboration} {TEXONO Collaboration}),\
  }\href {\doibase 10.1103/PhysRevD.81.072001} {\bibfield  {journal} {\bibinfo
  {journal} {Phys. Rev. D}\ }\textbf {\bibinfo {volume} {81}},\ \bibinfo
  {pages} {072001} (\bibinfo {year} {2010})}\BibitemShut {NoStop}%
\bibitem [{\citenamefont {Cheng}\ \emph {et~al.}(2004)\citenamefont {Cheng}
  \emph {et~al.}}]{Cheng2004}%
  \BibitemOpen
  \bibfield  {author} {\bibinfo {author} {\bibfnamefont {L.}~\bibnamefont
  {Cheng}} \emph {et~al.},\ }\href@noop {} {\emph {\bibinfo {title} {Physic and
  Safety Analysis for the NIST Research Reactor}}},\ \bibinfo {type} {Tech.
  Rep.}\ (\bibinfo  {institution} {National Institute of Standards and
  Technology},\ \bibinfo {year} {2004})\BibitemShut {NoStop}%
\bibitem [{\citenamefont {{D. Diamond, J. Baek, A. Hanson, L-Y. Cheng, N.
  Brown, and A. Cuadra}}(2014)}]{Diamond2014}%
  \BibitemOpen
  \bibfield  {author} {\bibinfo {author} {\bibnamefont {{D. Diamond, J. Baek,
  A. Hanson, L-Y. Cheng, N. Brown, and A. Cuadra}}},\ }\href@noop {} {\emph
  {\bibinfo {title} {{Conversion Preliminary Safety Analysis Report for the
  NIST Research Reactor}}}},\ \bibinfo {type} {Tech. Rep.}\ (\bibinfo
  {institution} {National Institute of Standards and Technology},\ \bibinfo
  {year} {2014})\ \bibinfo {note} {{BNL-107265-2015-IR}}\BibitemShut {NoStop}%
\bibitem [{\citenamefont {Stevens}\ \emph {et~al.}(2010)\citenamefont
  {Stevens}, \citenamefont {Tentner},\ and\ \citenamefont {Bergeron}}]{ILL}%
  \BibitemOpen
  \bibfield  {author} {\bibinfo {author} {\bibfnamefont {J.}~\bibnamefont
  {Stevens}}, \bibinfo {author} {\bibfnamefont {A.}~\bibnamefont {Tentner}}, \
  and\ \bibinfo {author} {\bibfnamefont {A.}~\bibnamefont {Bergeron}},\ }\href
  {\doibase 10.2172/986304} {\emph {\bibinfo {title} {Feasibility Analyses for
  HEU to LEU Fuel Cnversion of the LAUE Langivin Institute (ILL) High Flux
  Reactor (RHF)}}},\ \bibinfo {type} {Tech. Rep.}\ (\bibinfo  {institution}
  {Argonne National Laboratory},\ \bibinfo {year} {2010})\ \bibinfo {note}
  {aNL/RERTR/TM-10-21}\BibitemShut {NoStop}%
\bibitem [{\citenamefont {Koonen}(2004)}]{BR2}%
  \BibitemOpen
  \bibfield  {author} {\bibinfo {author} {\bibfnamefont {E.}~\bibnamefont
  {Koonen}},\ }\href@noop {} {\emph {\bibinfo {title} {Neutronic Modelling in
  Support of BR2 Irradiation Programmes}}},\ \bibinfo {type} {Tech. Rep.}\
  (\bibinfo  {institution} {Studiecentrum voor Kernenergie Centre d'Étude de
  l'énergie Nucléaire (SCK CEN)},\ \bibinfo {year} {2004})\BibitemShut
  {NoStop}%
\bibitem [{\citenamefont {Cheverton}\ and\ \citenamefont
  {Sims}(1971)}]{Cheverton}%
  \BibitemOpen
  \bibfield  {author} {\bibinfo {author} {\bibfnamefont {R.}~\bibnamefont
  {Cheverton}}\ and\ \bibinfo {author} {\bibfnamefont {T.}~\bibnamefont
  {Sims}},\ }\href@noop {} {\emph {\bibinfo {title} {{HFIR Core Nuclear
  Design}}}},\ \bibinfo {type} {Tech. Rep.}\ (\bibinfo  {institution} {Oak
  Ridge National Laboratory},\ \bibinfo {year} {1971})\BibitemShut {NoStop}%
\bibitem [{\citenamefont {Chandler}\ \emph {et~al.}(2016)\citenamefont
  {Chandler}, \citenamefont {Betzler}, \citenamefont {Hirtz}, \citenamefont
  {Ilas},\ and\ \citenamefont {Sunny}}]{Chandler2016}%
  \BibitemOpen
  \bibfield  {author} {\bibinfo {author} {\bibfnamefont {D.}~\bibnamefont
  {Chandler}}, \bibinfo {author} {\bibfnamefont {B.}~\bibnamefont {Betzler}},
  \bibinfo {author} {\bibfnamefont {G.}~\bibnamefont {Hirtz}}, \bibinfo
  {author} {\bibfnamefont {G.}~\bibnamefont {Ilas}}, \ and\ \bibinfo {author}
  {\bibfnamefont {E.}~\bibnamefont {Sunny}},\ }\href@noop {} {\emph {\bibinfo
  {title} {Modeling and Depletion Simulations for a High Flux Isotope Reactor
  Cycle with a Representative Experiment Loading}}},\ \bibinfo {type} {Tech.
  Rep.}\ (\bibinfo  {institution} {Oak Ridge National Laboratory},\ \bibinfo
  {year} {2016})\ \bibinfo {note} {{ORNL/TM-2016/23}}\BibitemShut {NoStop}%
\bibitem [{\citenamefont {Hogle}(2012)}]{Hogle2012}%
  \BibitemOpen
  \bibfield  {author} {\bibinfo {author} {\bibfnamefont {S.}~\bibnamefont
  {Hogle}},\ }\emph {\bibinfo {title} {Optimization of Transcurium Isotope
  Production in the High Flux Isotope Reactor}},\ \href@noop {} {Ph.D.
  thesis},\ \bibinfo  {school} {University of Tennessee, Knoxville} (\bibinfo
  {year} {2012})\BibitemShut {NoStop}%
\bibitem [{\citenamefont {Knight}\ \emph {et~al.}(1968)\citenamefont {Knight},
  \citenamefont {Binns},\ and\ \citenamefont {Adamson}}]{Knight1968}%
  \BibitemOpen
  \bibfield  {author} {\bibinfo {author} {\bibfnamefont {R.}~\bibnamefont
  {Knight}}, \bibinfo {author} {\bibfnamefont {J.}~\bibnamefont {Binns}}, \
  and\ \bibinfo {author} {\bibfnamefont {G.~J.}\ \bibnamefont {Adamson}},\
  }\href {\doibase 10.2172/4501160} {\emph {\bibinfo {title} {{Fabrication
  Procedures for Manufacturing High Flux Isotope Reactor Fuel Elements}}}},\
  \bibinfo {type} {Tech. Rep.}\ (\bibinfo  {institution} {Oak Ridge National
  Laboratory},\ \bibinfo {year} {1968})\ \bibinfo {note}
  {oRNL-4242}\BibitemShut {NoStop}%
\bibitem [{\citenamefont {Chandler}\ and\ \citenamefont
  {Ellis}(2015)}]{Chandler2015}%
  \BibitemOpen
  \bibfield  {author} {\bibinfo {author} {\bibfnamefont {D.}~\bibnamefont
  {Chandler}}\ and\ \bibinfo {author} {\bibfnamefont {R.}~\bibnamefont
  {Ellis}},\ }\href@noop {} {\bibfield  {journal} {\bibinfo  {journal}
  {Proceedings of the Nuclear and Emerging Technologies for Space (NETS)
  Conference}\ } (\bibinfo {year} {2015})}\BibitemShut {NoStop}%
\bibitem [{\citenamefont {Chandler}(2016)}]{Chandler2016physor}%
  \BibitemOpen
  \bibfield  {author} {\bibinfo {author} {\bibfnamefont {D.}~\bibnamefont
  {Chandler}},\ }\href@noop {} {\bibfield  {journal} {\bibinfo  {journal}
  {PHYSOR Conference}\ } (\bibinfo {year} {2016})}\BibitemShut {NoStop}%
\bibitem [{\citenamefont {Hurt}\ \emph {et~al.}(2016)\citenamefont {Hurt},
  \citenamefont {Freels}, \citenamefont {Hobbs}, \citenamefont {Jain},\ and\
  \citenamefont {Maldonado}}]{Hurt2016}%
  \BibitemOpen
  \bibfield  {author} {\bibinfo {author} {\bibfnamefont {C.}~\bibnamefont
  {Hurt}}, \bibinfo {author} {\bibfnamefont {J.}~\bibnamefont {Freels}},
  \bibinfo {author} {\bibfnamefont {R.}~\bibnamefont {Hobbs}}, \bibinfo
  {author} {\bibfnamefont {P.}~\bibnamefont {Jain}}, \ and\ \bibinfo {author}
  {\bibfnamefont {G.}~\bibnamefont {Maldonado}},\ }\href@noop {} {\emph
  {\bibinfo {title} {{Thermal Safety Analysis for the Production of
  Plutonium-238 at the High Flux Isotope Reactor}}}},\ \bibinfo {type} {Tech.
  Rep.}\ (\bibinfo  {institution} {Oak Ridge National Laboratory},\ \bibinfo
  {year} {2016})\ \bibinfo {note} {{ORNL/TM-2016/234}}\BibitemShut {NoStop}%
\bibitem [{\citenamefont {{X-5 Monte Carlo Team}}(2005)}]{MCNP5}%
  \BibitemOpen
  \bibfield  {author} {\bibinfo {author} {\bibnamefont {{X-5 Monte Carlo
  Team}}},\ }\href@noop {} {\emph {\bibinfo {title} {{MCNP---A General Monte
  Carlo N-Particle Transport Code, Version 5}}}},\ \bibinfo {type} {Tech.
  Rep.}\ (\bibinfo  {institution} {Los Alamos National Laboratory},\ \bibinfo
  {year} {2005})\ \bibinfo {note} {{LA-UR-03-1987}}\BibitemShut {NoStop}%
\bibitem [{\citenamefont {Werner}(2017)}]{MCNP6}%
  \BibitemOpen
  \bibfield  {author} {\bibinfo {author} {\bibfnamefont {C.}~\bibnamefont
  {Werner}},\ }\href@noop {} {\emph {\bibinfo {title} {MCNP User's
  Manual---Code Version~6.2}}},\ \bibinfo {organization} {Los Alamos National
  Laboratory} (\bibinfo {year} {2017})\BibitemShut {NoStop}%
\bibitem [{\citenamefont {Xoubi}\ \emph {et~al.}(2004)\citenamefont {Xoubi},
  \citenamefont {{Primm III}} \emph {et~al.}}]{Xoubi2004}%
  \BibitemOpen
  \bibfield  {author} {\bibinfo {author} {\bibfnamefont {N.}~\bibnamefont
  {Xoubi}}, \bibinfo {author} {\bibfnamefont {T.}~\bibnamefont {{Primm III}}},
  \emph {et~al.},\ }\href@noop {} {\emph {\bibinfo {title} {Modeling of the
  High Flux Isotope Reactor Cycle 400}}},\ \bibinfo {type} {Tech. Rep.}\
  (\bibinfo  {institution} {Oak Ridge National Laboratory},\ \bibinfo {year}
  {2004})\ \bibinfo {note} {{ORNL/TM-2004/251}}\BibitemShut {NoStop}%
\bibitem [{\citenamefont {Ilas}\ \emph {et~al.}(2015)\citenamefont {Ilas},
  \citenamefont {Chandler} \emph {et~al.}}]{Ilas2015}%
  \BibitemOpen
  \bibfield  {author} {\bibinfo {author} {\bibfnamefont {G.}~\bibnamefont
  {Ilas}}, \bibinfo {author} {\bibfnamefont {D.}~\bibnamefont {Chandler}},
  \emph {et~al.},\ }\href@noop {} {\emph {\bibinfo {title} {Modeling and
  Simulations for the High Flux Isotope Reactor Cycle 400}}},\ \bibinfo {type}
  {Tech. Rep.}\ (\bibinfo  {institution} {Oak Ridge National Laboratory},\
  \bibinfo {year} {2015})\ \bibinfo {note} {oRNL/TM-2015/36}\BibitemShut
  {NoStop}%
\bibitem [{\citenamefont {Haeck}(2012)}]{VESTA}%
  \BibitemOpen
  \bibfield  {author} {\bibinfo {author} {\bibfnamefont {W.}~\bibnamefont
  {Haeck}},\ }\href@noop {} {\emph {\bibinfo {title} {VESTA User's
  Manual---Version~2.1.0}}},\ \bibinfo {organization} {Institut de
  Radioprotection et de Surete Nucleaire, France} (\bibinfo {year}
  {2012})\BibitemShut {NoStop}%
\bibitem [{\citenamefont {{L. Snoj and M. Ravnik}}(2006)}]{Snoj2006}%
  \BibitemOpen
  \bibfield  {author} {\bibinfo {author} {\bibnamefont {{L. Snoj and M.
  Ravnik}}},\ }\href@noop {} {\bibfield  {journal} {\bibinfo  {journal}
  {Proceedings of the International Conference Nuclear Energy for New Europe
  2006}\ } (\bibinfo {year} {2006})}\BibitemShut {NoStop}%
\bibitem [{\citenamefont {Sterbentz}(2013)}]{Sterbentz2013}%
  \BibitemOpen
  \bibfield  {author} {\bibinfo {author} {\bibfnamefont {J.}~\bibnamefont
  {Sterbentz}},\ }\href@noop {} {\emph {\bibinfo {title} {Q-value (MeV/fission)
  Determination for the Advanced Test Reactor}}},\ \bibinfo {type} {Tech.
  Rep.}\ (\bibinfo  {institution} {Idaho National Laboratory},\ \bibinfo {year}
  {2013})\ \bibinfo {note} {iNL/EXT-13-29256}\BibitemShut {NoStop}%
\bibitem [{\citenamefont {Kopeikin}(2012)}]{Kopeikin2012}%
  \BibitemOpen
  \bibfield  {author} {\bibinfo {author} {\bibfnamefont {V.}~\bibnamefont
  {Kopeikin}},\ }\href@noop {} {\bibfield  {journal} {\bibinfo  {journal}
  {Phys. of At. Nuclei}\ }\textbf {\bibinfo {volume} {75}} (\bibinfo {year}
  {2012})}\BibitemShut {NoStop}%
\bibitem [{\citenamefont {Ma}\ \emph {et~al.}(2013)\citenamefont {Ma},
  \citenamefont {Zhong}, \citenamefont {Wang}, \citenamefont {Chen},\ and\
  \citenamefont {Cao}}]{Ma2013}%
  \BibitemOpen
  \bibfield  {author} {\bibinfo {author} {\bibfnamefont {X.~B.}\ \bibnamefont
  {Ma}}, \bibinfo {author} {\bibfnamefont {W.~L.}\ \bibnamefont {Zhong}},
  \bibinfo {author} {\bibfnamefont {L.~Z.}\ \bibnamefont {Wang}}, \bibinfo
  {author} {\bibfnamefont {Y.~X.}\ \bibnamefont {Chen}}, \ and\ \bibinfo
  {author} {\bibfnamefont {J.}~\bibnamefont {Cao}},\ }\href {\doibase
  10.1103/PhysRevC.88.014605} {\bibfield  {journal} {\bibinfo  {journal} {Phys.
  Rev. C}\ }\textbf {\bibinfo {volume} {88}},\ \bibinfo {pages} {014605}
  (\bibinfo {year} {2013})}\BibitemShut {NoStop}%
\bibitem [{\citenamefont {Zerovnik}\ \emph {et~al.}(2014)\citenamefont
  {Zerovnik} \emph {et~al.}}]{Zerovnik2014}%
  \BibitemOpen
  \bibfield  {author} {\bibinfo {author} {\bibfnamefont {G.}~\bibnamefont
  {Zerovnik}} \emph {et~al.},\ }\href@noop {} {\bibfield  {journal} {\bibinfo
  {journal} {Ann. of Nucl. Energy}\ }\textbf {\bibinfo {volume} {63}},\
  \bibinfo {pages} {126} (\bibinfo {year} {2014})}\BibitemShut {NoStop}%
\bibitem [{\citenamefont {Rearden}\ \emph {et~al.}(2011)\citenamefont
  {Rearden}, \citenamefont {Jessee} \emph {et~al.}}]{SCALE62}%
  \BibitemOpen
  \bibfield  {author} {\bibinfo {author} {\bibfnamefont {B.}~\bibnamefont
  {Rearden}}, \bibinfo {author} {\bibfnamefont {M.}~\bibnamefont {Jessee}},
  \emph {et~al.},\ }\href@noop {} {\emph {\bibinfo {title} {{SCALE: A
  Comprehensive Modeling and Simulation Suite for Nuclear Safety Analysis and
  Design}}}},\ \bibinfo {type} {Tech. Rep.}\ (\bibinfo  {institution} {Oak
  Ridge National Laboratory},\ \bibinfo {year} {2011})\ \bibinfo {note}
  {{ORNL/TM-2005/39}}\BibitemShut {NoStop}%
\bibitem [{\citenamefont {Peplow}(2004)}]{Peplow2004}%
  \BibitemOpen
  \bibfield  {author} {\bibinfo {author} {\bibfnamefont {D.}~\bibnamefont
  {Peplow}},\ }\href@noop {} {\emph {\bibinfo {title} {A Computational Model of
  the High Flux Isotope Reactor for the Calculation of Cold Source, Beam Tube,
  and Guide Hall Nuclear Parameters}}},\ \bibinfo {type} {Tech. Rep.}\
  (\bibinfo  {institution} {Oak Ridge National Laboratory},\ \bibinfo {year}
  {2004})\ \bibinfo {note} {{ORNL/TM-2004/237}}\BibitemShut {NoStop}%
\bibitem [{\citenamefont {Díez}\ \emph {et~al.}(2015)\citenamefont {Díez},
  \citenamefont {Buss}, \citenamefont {Hoefer}, \citenamefont {Porsch},\ and\
  \citenamefont {Cabellos}}]{Diez2015}%
  \BibitemOpen
  \bibfield  {author} {\bibinfo {author} {\bibfnamefont {C.}~\bibnamefont
  {Díez}}, \bibinfo {author} {\bibfnamefont {O.}~\bibnamefont {Buss}},
  \bibinfo {author} {\bibfnamefont {A.}~\bibnamefont {Hoefer}}, \bibinfo
  {author} {\bibfnamefont {D.}~\bibnamefont {Porsch}}, \ and\ \bibinfo {author}
  {\bibfnamefont {O.}~\bibnamefont {Cabellos}},\ }\href {\doibase
  https://doi.org/10.1016/j.anucene.2014.10.022} {\bibfield  {journal}
  {\bibinfo  {journal} {Ann. of Nucl. Energy}\ }\textbf {\bibinfo {volume}
  {77}},\ \bibinfo {pages} {101 } (\bibinfo {year} {2015})}\BibitemShut
  {NoStop}%
\bibitem [{\citenamefont {Mosteller}(2010)}]{MCNPvalidation}%
  \BibitemOpen
  \bibfield  {author} {\bibinfo {author} {\bibfnamefont {R.}~\bibnamefont
  {Mosteller}},\ }\href@noop {} {\emph {\bibinfo {title} {{An Expanded
  Criticality Validation Suite for MCNP}}}},\ \bibinfo {type} {Tech. Rep.}\
  (\bibinfo  {institution} {Los Alamos National Laboratory},\ \bibinfo {year}
  {2010})\ \bibinfo {note} {{LA-UR-10-06230}}\BibitemShut {NoStop}%
\bibitem [{\citenamefont {Ilas}\ \emph {et~al.}(2010)\citenamefont {Ilas},
  \citenamefont {Gauld}, \citenamefont {Difilippo},\ and\ \citenamefont
  {Emmett}}]{Ilas2010}%
  \BibitemOpen
  \bibfield  {author} {\bibinfo {author} {\bibfnamefont {G.}~\bibnamefont
  {Ilas}}, \bibinfo {author} {\bibfnamefont {I.}~\bibnamefont {Gauld}},
  \bibinfo {author} {\bibfnamefont {F.}~\bibnamefont {Difilippo}}, \ and\
  \bibinfo {author} {\bibfnamefont {M.}~\bibnamefont {Emmett}},\ }\href@noop {}
  {\emph {\bibinfo {title} {Analysis of Experimental Data for High Burnup PWR
  Spent Fuel Isotopic Validation: Calvert Cliffs, Takahama, and Three Mile
  Island Reactors}}},\ \bibinfo {type} {Tech. Rep.}\ (\bibinfo  {institution}
  {Oak Ridge National Laboratory},\ \bibinfo {year} {2010})\ \bibinfo {note}
  {{ORNL/TM-2008/071}}\BibitemShut {NoStop}%
\bibitem [{\citenamefont {Conant}(2019)}]{Conant2019}%
  \BibitemOpen
  \bibfield  {author} {\bibinfo {author} {\bibfnamefont {A.}~\bibnamefont
  {Conant}},\ }\emph {\bibinfo {title} {Antineutrino Spectrum Characterization
  of the High Flux Isotope Reactor Using Neutronic Simulations}},\ \href@noop
  {} {Ph.D. thesis},\ \bibinfo  {school} {Georgia Institute of Technology}
  (\bibinfo {year} {2019})\BibitemShut {NoStop}%
\bibitem [{\citenamefont {Radaideh}\ \emph {et~al.}(2018)\citenamefont
  {Radaideh}, \citenamefont {Price},\ and\ \citenamefont
  {Kozlowski}}]{Radaideh2018}%
  \BibitemOpen
  \bibfield  {author} {\bibinfo {author} {\bibfnamefont {M.}~\bibnamefont
  {Radaideh}}, \bibinfo {author} {\bibfnamefont {D.}~\bibnamefont {Price}}, \
  and\ \bibinfo {author} {\bibfnamefont {T.}~\bibnamefont {Kozlowski}},\
  }\href@noop {} {\bibfield  {journal} {\bibinfo  {journal} {Proceedings of The
  Fourth International Conference on Physics and Technology of Reactors and
  Applications (PHYTRA4)}\ } (\bibinfo {year} {2018})}\BibitemShut {NoStop}%
\bibitem [{\citenamefont {Dwyer}()}]{Oklo}%
  \BibitemOpen
  \bibfield  {author} {\bibinfo {author} {\bibfnamefont {D.}~\bibnamefont
  {Dwyer}},\ }\href@noop {} {\enquote {\bibinfo {title} {{OKLO: A toolkit for
  Modeling Nuclides and Nuclear Reactions}},}\ }\bibinfo {howpublished}
  {\url{https://github.com/dadwyer/oklo}}\BibitemShut {NoStop}%
\bibitem [{\citenamefont {Wilkinson}(1989)}]{Wilkinson1989}%
  \BibitemOpen
  \bibfield  {author} {\bibinfo {author} {\bibfnamefont {D.}~\bibnamefont
  {Wilkinson}},\ }\href {\doibase https://doi.org/10.1016/0168-9002(89)90712-2}
  {\bibfield  {journal} {\bibinfo  {journal} {Nucl. Instrum. and Methods in
  Phys. Res. Sect. A}\ }\textbf {\bibinfo {volume} {275}},\ \bibinfo {pages}
  {378 } (\bibinfo {year} {1989})}\BibitemShut {NoStop}%
\bibitem [{\citenamefont {Schenter}\ and\ \citenamefont
  {Vogel}(1983)}]{Schenter1983}%
  \BibitemOpen
  \bibfield  {author} {\bibinfo {author} {\bibfnamefont {G.~K.}\ \bibnamefont
  {Schenter}}\ and\ \bibinfo {author} {\bibfnamefont {P.}~\bibnamefont
  {Vogel}},\ }\href {\doibase 10.13182/NSE83-A17574} {\bibfield  {journal}
  {\bibinfo  {journal} {Nucl. Sci. and Eng.}\ }\textbf {\bibinfo {volume}
  {83}},\ \bibinfo {pages} {393} (\bibinfo {year} {1983})}\BibitemShut
  {NoStop}%
\bibitem [{\citenamefont {Huber}(2016)}]{Huber2016}%
  \BibitemOpen
  \bibfield  {author} {\bibinfo {author} {\bibfnamefont {P.}~\bibnamefont
  {Huber}},\ }\href {\doibase https://doi.org/10.1016/j.nuclphysb.2016.04.012}
  {\bibfield  {journal} {\bibinfo  {journal} {Nucl. Phys. B}\ }\textbf
  {\bibinfo {volume} {908}},\ \bibinfo {pages} {268 } (\bibinfo {year}
  {2016})}\BibitemShut {NoStop}%
\bibitem [{\citenamefont {Schmidt}\ \emph {et~al.}(1982)\citenamefont {Schmidt}
  \emph {et~al.}}]{Al28}%
  \BibitemOpen
  \bibfield  {author} {\bibinfo {author} {\bibfnamefont {H.}~\bibnamefont
  {Schmidt}} \emph {et~al.},\ }\href {\doibase 10.1103/PhysRevC.25.2888}
  {\bibfield  {journal} {\bibinfo  {journal} {Phys. Rev. C}\ }\textbf {\bibinfo
  {volume} {25}} (\bibinfo {year} {1982}),\
  10.1103/PhysRevC.25.2888}\BibitemShut {NoStop}%
\bibitem [{\citenamefont {Conant}\ \emph {et~al.}(2018)\citenamefont {Conant},
  \citenamefont {Mumm},\ and\ \citenamefont {Erickson}}]{Conant2018physor}%
  \BibitemOpen
  \bibfield  {author} {\bibinfo {author} {\bibfnamefont {A.}~\bibnamefont
  {Conant}}, \bibinfo {author} {\bibfnamefont {H.}~\bibnamefont {Mumm}}, \ and\
  \bibinfo {author} {\bibfnamefont {A.}~\bibnamefont {Erickson}},\ }\href@noop
  {} {\bibfield  {journal} {\bibinfo  {journal} {PHYSOR Conference}\ }
  (\bibinfo {year} {2018})}\BibitemShut {NoStop}%
\bibitem [{\citenamefont {Junde}(2008)}]{Junde2008}%
  \BibitemOpen
  \bibfield  {author} {\bibinfo {author} {\bibfnamefont {H.}~\bibnamefont
  {Junde}},\ }\href {\doibase https://doi.org/10.1016/j.nds.2008.03.001}
  {\bibfield  {journal} {\bibinfo  {journal} {Nuclear Data Sheets}\ }\textbf
  {\bibinfo {volume} {109}},\ \bibinfo {pages} {787 } (\bibinfo {year}
  {2008})}\BibitemShut {NoStop}%
\bibitem [{\citenamefont {Browne}\ and\ \citenamefont
  {Tuli}(2010)}]{Browne2010}%
  \BibitemOpen
  \bibfield  {author} {\bibinfo {author} {\bibfnamefont {E.}~\bibnamefont
  {Browne}}\ and\ \bibinfo {author} {\bibfnamefont {J.}~\bibnamefont {Tuli}},\
  }\href {\doibase https://doi.org/10.1016/j.nds.2010.03.004} {\bibfield
  {journal} {\bibinfo  {journal} {Nuclear Data Sheets}\ }\textbf {\bibinfo
  {volume} {111}},\ \bibinfo {pages} {1093 } (\bibinfo {year}
  {2010})}\BibitemShut {NoStop}%
\bibitem [{\citenamefont {Junde}\ \emph {et~al.}(2011)\citenamefont {Junde},
  \citenamefont {Su},\ and\ \citenamefont {Dong}}]{Junde2011}%
  \BibitemOpen
  \bibfield  {author} {\bibinfo {author} {\bibfnamefont {H.}~\bibnamefont
  {Junde}}, \bibinfo {author} {\bibfnamefont {H.}~\bibnamefont {Su}}, \ and\
  \bibinfo {author} {\bibfnamefont {Y.}~\bibnamefont {Dong}},\ }\href {\doibase
  https://doi.org/10.1016/j.nds.2011.04.004} {\bibfield  {journal} {\bibinfo
  {journal} {Nuclear Data Sheets}\ }\textbf {\bibinfo {volume} {112}},\
  \bibinfo {pages} {1513 } (\bibinfo {year} {2011})}\BibitemShut {NoStop}%
\bibitem [{\citenamefont {Tilley}\ \emph {et~al.}(2002)\citenamefont {Tilley},
  \citenamefont {Cheves}, \citenamefont {Godwin}, \citenamefont {Hale},
  \citenamefont {Hofmann}, \citenamefont {Kelley}, \citenamefont {Sheu},\ and\
  \citenamefont {Weller}}]{Tilley2003}%
  \BibitemOpen
  \bibfield  {author} {\bibinfo {author} {\bibfnamefont {D.}~\bibnamefont
  {Tilley}}, \bibinfo {author} {\bibfnamefont {C.}~\bibnamefont {Cheves}},
  \bibinfo {author} {\bibfnamefont {J.}~\bibnamefont {Godwin}}, \bibinfo
  {author} {\bibfnamefont {G.}~\bibnamefont {Hale}}, \bibinfo {author}
  {\bibfnamefont {H.}~\bibnamefont {Hofmann}}, \bibinfo {author} {\bibfnamefont
  {J.}~\bibnamefont {Kelley}}, \bibinfo {author} {\bibfnamefont
  {C.}~\bibnamefont {Sheu}}, \ and\ \bibinfo {author} {\bibfnamefont
  {H.}~\bibnamefont {Weller}},\ }\href {\doibase
  https://doi.org/10.1016/S0375-9474(02)00597-3} {\bibfield  {journal}
  {\bibinfo  {journal} {Nucl. Phys. A}\ }\textbf {\bibinfo {volume} {708}},\
  \bibinfo {pages} {3 } (\bibinfo {year} {2002})}\BibitemShut {NoStop}%
\bibitem [{\citenamefont {Dong}\ and\ \citenamefont {Junde}(2015)}]{Dong2015}%
  \BibitemOpen
  \bibfield  {author} {\bibinfo {author} {\bibfnamefont {Y.}~\bibnamefont
  {Dong}}\ and\ \bibinfo {author} {\bibfnamefont {H.}~\bibnamefont {Junde}},\
  }\href {\doibase https://doi.org/10.1016/j.nds.2015.08.003} {\bibfield
  {journal} {\bibinfo  {journal} {Nuclear Data Sheets}\ }\textbf {\bibinfo
  {volume} {128}},\ \bibinfo {pages} {185 } (\bibinfo {year}
  {2015})}\BibitemShut {NoStop}%
\bibitem [{\citenamefont {Nica}(2010)}]{Nica2015}%
  \BibitemOpen
  \bibfield  {author} {\bibinfo {author} {\bibfnamefont {N.}~\bibnamefont
  {Nica}},\ }\href {\doibase https://doi.org/10.1016/j.nds.2010.03.001}
  {\bibfield  {journal} {\bibinfo  {journal} {Nuclear Data Sheets}\ }\textbf
  {\bibinfo {volume} {111}},\ \bibinfo {pages} {525 } (\bibinfo {year}
  {2010})}\BibitemShut {NoStop}%
\bibitem [{\citenamefont {Zinkle}\ and\ \citenamefont
  {Busby}(2009)}]{Zinkle2009}%
  \BibitemOpen
  \bibfield  {author} {\bibinfo {author} {\bibfnamefont {S.~J.}\ \bibnamefont
  {Zinkle}}\ and\ \bibinfo {author} {\bibfnamefont {J.~T.}\ \bibnamefont
  {Busby}},\ }\href {\doibase https://doi.org/10.1016/S1369-7021(09)70294-9}
  {\bibfield  {journal} {\bibinfo  {journal} {Mater. Today}\ }\textbf {\bibinfo
  {volume} {12}},\ \bibinfo {pages} {12 } (\bibinfo {year} {2009})}\BibitemShut
  {NoStop}%
\bibitem [{\citenamefont {{McConn Jr.}}\ \emph {et~al.}(2011)\citenamefont
  {{McConn Jr.}} \emph {et~al.}}]{PNNLcompendium}%
  \BibitemOpen
  \bibfield  {author} {\bibinfo {author} {\bibfnamefont {R.}~\bibnamefont
  {{McConn Jr.}}} \emph {et~al.},\ }\href@noop {} {\emph {\bibinfo {title}
  {{Compendium of Material Composition Data for Radiation Transport
  Modeling}}}},\ \bibinfo {type} {Tech. Rep.}\ (\bibinfo  {institution}
  {Pacific Northwest National Laboratory},\ \bibinfo {year} {2011})\ \bibinfo
  {note} {pIET-43741-TM-963}\BibitemShut {NoStop}%
\bibitem [{\citenamefont {Ashenfelter}\ \emph {et~al.}(2019)\citenamefont
  {Ashenfelter} \emph {et~al.}}]{PROSPECTprl2019}%
  \BibitemOpen
  \bibfield  {author} {\bibinfo {author} {\bibfnamefont {J.}~\bibnamefont
  {Ashenfelter}} \emph {et~al.} (\bibinfo {collaboration} {PROSPECT
  Collaboration}),\ }\href {\doibase 10.1103/PhysRevLett.122.251801} {\bibfield
   {journal} {\bibinfo  {journal} {Phys. Rev. Lett.}\ }\textbf {\bibinfo
  {volume} {122}},\ \bibinfo {pages} {251801} (\bibinfo {year}
  {2019})}\BibitemShut {NoStop}%
\end{thebibliography}%
